\documentclass[runningheads]{svmult}

\usepackage{makeidx}   
\usepackage{graphicx}  
\usepackage{subeqnar}  
\usepackage{multicol}  
\usepackage{physprbb}  
\makeindex             

\usepackage{epsfig}
\usepackage{url}
\usepackage{citesort}
\usepackage{amssymb}

\newcommand{\vx}{{\vec x}}
\newcommand{\sh}{\Sigma_{\cH}}

\newcommand{\vh}{{\vec h}}

\newcommand{\vy}{{\vec y}}
\newcommand{\vz}{{\vec z}}
\newcommand{\vg}{\gamma}
\newcommand{\dG}{\dot \Gamma}
\newcommand{\tgs}{{t(s)}}
\newcommand{\pxhere}{{\phi_{\vx}(s)}}
\newcommand{\pxpm}{{\phi_{\pm}(s)}}

\newcommand{\wnc}{weakly null convex}
\newcommand{\id}{{\textrm{id}}}
\newcommand{\mycalu}{{\mycal U}}
\newcommand{\mcU}{{\mycal U}}
\newcommand{\mcV}{{\mycal V}}
\newcommand{\mcC}{{\mycal C}}
\newcommand{\mcT}{{\mycal T}}
\newcommand{\mcH}{{\mycal H}}
\newcommand{\mycals}{{\mycal S}}
\newcommand{\mycalm}{{\mcM}}
\newcommand{\mcE}{{\mycal E}}

\newcommand{\pihyp}{\partial_\infty\hyp}
\newcommand{\cMext}{{\mcM}_{\ext}}
\newcommand{\timkd}{{\mycal T}^\flat}
\newcommand{\timku}{{\mycal T}^\sharp}

\newcommand{\cimkd}{{\mycal C}^\flat}
\newcommand{\cimku}{{\mycal C}^\sharp}
\newcommand{\proj}{\textrm{pr}}
\newcommand{\mcM}{{\mycal M}}
\newcommand{\mcO}{{\mycal O}}
\newcommand{\Mclosed}{\bcM}
\newcommand{\Sect}[1]{Sect.~\ref{#1}}
\newcommand{\cg}{\,{\tilde {\!g}}}

\newcommand{\event}{{\mycal E}}

\newcommand{\Hess}{{\textrm{Hess}}}

\newcommand{\cA}{{\mycal A}}

\ifx\macrosloaded\relax\endinput\else\let\macrosloaded\relax\fi

\newcommand{\tA}{\theta_\Al}

\newcommand{\Al}{{\mathcal Al}}

\newcommand{\BA}{B_\Al}

\newcommand{\cS}{{\cal S}}

\newcommand{\bel}[1]{\begin{equation}\label{#1}}

\newcommand{\cP}{{\cal P}}

\newcommand{\dg}{\dot\gamma}

\newcommand{\ext}{{\mathrm{ext}}}

\newcommand{\bcM}{{\,\,\,\overline{\!\!\!\mcM}}}
\newcommand{\tcM}{\,\,\,\,\widetilde{\!\!\!\!\cM}}
\newcommand{\tg}{{\tilde g}}

\newcommand{\be}{\begin{equation}}
\newcommand{\bea}{\begin{eqnarray}}
\newcommand{\eea}{\end{eqnarray}}
\newcommand{\beaa}{\begin{eqnarray*}}
\newcommand{\eeaa}{\end{eqnarray*}}
\newcommand{\bseq}{\begin{subeq}}
\newcommand{\eseq}{\end{subeq}}

\def \rectangle#1#2{\hbox{\vrule\vbox to #2
{\hrule\hbox to
#1{\hfil}\vfil\hrule}\vrule}}\def\square{\,\,\rectangle{7pt}{7
pt}\,\,}
\newcommand{\edd}{\end{document}}

\newcommand{\ro}{\rho}

\newcommand{\Ric}{\mbox{Ric}}

\def\frac#1#2{{{#1}\over{#2}}}

\newcommand{\hyp}{{\mycal S}}

\newcommand{\eeq}{\end{equation}}
\newcommand{\ee}{\end{equation}}
\newcommand{\beqa}{\begin{eqnarray}}
\newcommand{\beqar}{\begin{deqarr}}
\newcommand{\eeqa}{\end{eqnarray}}
\newcommand{\eeqar}{\end{deqarr}}
\newcommand{\beqan}{\begin{eqnarray*}}
\newcommand{\eeqan}{\end{eqnarray*}}
\newcommand{\ba}{\begin{array}}
\newcommand{\ea}{\end{array}}
\newcommand{\mcB}{{\mycal B}}

\newcommand{\cH}{{\mycal H}}
\newcommand{\cB}{{\mycal B}}

\newcommand{\cU}{{\cal U}}
\newcommand{\const}{\mbox{\rm const}} 

\DeclareFontFamily{OT1}{rsfs}{}
\DeclareFontShape{OT1}{rsfs}{m}{n}{ <-7> rsfs5 <7-10> rsfs7 <10->
rsfs10}{} \DeclareMathAlphabet{\mycal}{OT1}{rsfs}{m}{n}
\def\scri{{\mycal I}}%
\def\scrip{\scri^{+}}%
\def\Scri{\scri}

\newcounter{mnotecount}[section]

\newcommand{\R}{\mathbb R}

\newcommand{\eq}[1]{(\ref{#1})}

\newcommand{\Eqs}[2]{\eq{#1}-\eq{#2}} 
\newcommand{\eqs}[2]{\eq{#1}-\eq{#2}}

\newcommand{\bmcM}{\,\,\,\,\overline{\!\!\!\!\mycal M}}

\newcommand{\cM}{\mycal M}

\begin{document}\newcommand{\chnindex}[1]{\index{#1}}
\newcommand{\chindex}[1]{\index{#1}}
\newcommand{\bhindex}[1]{\chindex{black holes!#1}}
\newtheorem{ptcTheorem} {Theorem}
\newtheorem{ptctheorem} [ptcTheorem]{Theorem}
\newtheorem{ptcCorollary} [ptcTheorem] {Corollary}
\newtheorem{ptcLemma} [ptcTheorem] {Lemma}
\newtheorem{ptcProposition} [ptcTheorem] {Proposition}
\newtheorem{ptcproposition} [ptcTheorem] {Proposition}
\newtheorem{ptcConjecture}[ptcTheorem]{Conjecture}
\newtheorem{aptctheorem} {Theorem}
\newtheorem{aptcproposition} [aptctheorem] {Proposition}
\newtheorem{aptccorollary} [aptctheorem] {Corollary}
\renewcommand{\theaptctheorem}{A.\arabic{aptctheorem}}
\title{Black holes}
\toctitle{Black holes}
\titlerunning{Black holes}

\author{Piotr T. Chru\'sciel\thanks{Partially supported by a Polish Research Council grant \# KBN 2 P03B 130
16.}}
\authorrunning{P.~T.~Chru\'sciel}
\institute{
 D\'epartement de Math\'ematiques, Facult\'e des Sciences,
Parc de Grandmont, \\ F 37200 Tours, France}

\maketitle              
\chindex{black holes|(}
\begin{abstract}
This paper is concerned with several not-quantum aspects of black
holes, with emphasis on theoretical and mathematical issues
related to numerical modeling of black hole space-times. Part of
the material has a review character, but some new results or
proposals are also presented. We review the experimental evidence
for existence of black holes. We propose a definition of black
hole region for any theory governed by a symmetric hyperbolic
system of equations. Our definition reproduces the usual one for
gravity, and leads to the one associated with the Unruh metric in
the case of Euler equations. We review the global conditions which
have been used in the Scri-based definition of a black hole and
point out the deficiencies of the Scri  approach. Various results
on the structure of horizons and apparent horizons are presented,
and a new proof of semi-convexity of horizons based on a
variational principle is given. Recent results on the
classification of stationary singularity-free vacuum solutions are
reviewed. Two new frameworks for discussing black holes are
proposed: a ``naive approach", based on  coordinate systems, and a
``quasi-local approach", based on timelike boundaries satisfying a
null convexity condition. Some properties of the resulting black
holes are established, including an area theorem, topology
theorems, and an approximation theorem for the location of the
horizon.
\end{abstract}\section{Introduction}

Black holes belong to the most fascinating objects predicted by
Einstein's theory of gravitation.  Although they have been studied
for years,\footnote{The reader is referred to the introduction
to~\cite{CMSciama} for an excellent concise review of the history
of the concept of a black hole, and
to~\cite{Carter:1997im,Israel:bhreview} for more detailed ones.}
they still attract quite a lot of attention in the physics and
astrophysics literature and, in fact, an exponential growth of the
number of related papers can be observed.\footnote{A search on the
key word ``black holes" on 2.XI.2001 reveals 213 papers on gr-qc
for the current year and 773 papers from 1991, date of the
beginning of the archive; the figures on astro-ph and hep-th are
very similar.}
%
It turns out that several field theories are known to possess
solutions which exhibit black hole properties:
\begin{itemize}
\item The ``standard" gravitational ones which, according to our
  current postulates, are black holes for all classical fields.
\item The ``dumb holes", which are the sonic counterparts of black
  holes, first discussed by Unruh~\cite{Unruh}. \item The ``optical"
  ones -- the black-hole counterparts arising in the theory of moving
  dielectric media, or in non-linear electrodynamics
  \cite{Leonhardt:Piwnicki,Novello:2001fv}.
\item The ``numerical black holes" -- objects constructed by numerical
  general relativists. We shall argue below that this leads to the
  need of introducing, and studying, new frameworks for the notion of
  black holes; two such frameworks (``naive black holes" and
  ``quasi-local black holes") will be introduced and studied in
  Sections~\ref{Snbh} and \ref{Sqlbh} below.
\end{itemize}
(An even longer list of models and submodels can be found
in~\cite{Barcelo:2001ah}.) In this work we shall discuss various
aspects of the above. The reader is referred to
\cite{Jacobson:1999zk,Brout:1995rd,Wald:LR,Horowitz:1996rn,Peet:2000hn,Das:2000su}
and references therein for a review of quantum aspects of black
holes. Let us start with a short review of the observational
status of black holes in astrophysics.

\section{Experimental evidence}
\bhindex{gravitational!astrophysical|(} While there is growing
evidence that black holes do indeed exist in astrophysical
objects, and that alternative explanations for the observations
discussed below seem less convincing, it should be borne in mind
that no undisputed evidence of occurrence of black holes has been
presented so far. The flagship black hole candidate used to be
Cygnus X-1, known and studied for years ({\em cf.,
e.g.,}\/~\cite{CMSciama}). Its published mass has been going up
and down over time: an optimistic interpretation of this
phenomenon is that there has been mass accretion during the
ascending periods, and Hawking radiation during the descending
ones; a more realistic one is that there is still considerable
uncertainty in the determination of this mass.
Table~\ref{luminet:B}\footnote{The recent
  review~\cite{NGM} lists thirteen black hole candidates.} lists a
series of very strong black hole candidates in $X$-ray binary
systems; $M_c$ is mass of the compact object and $M_*$ is that of
its optical companion; some other candidates, as well as
references, can be found in~\cite{Narayan,NGM}. The binaries have
been divided into two families: the High Mass X-ray Binaries
(HMXB), where the companion star is of (relatively) high mass, and
the Low Mass X-ray Binaries (LMXB), where the companion is
typically below a solar mass. The LMXB's include the "X-ray
transients", so-called because of flaring-up behaviour. This
particularity allows to make detailed studies of their optical
properties during the quiescent periods, which would be impossible
during the periods of intense $X$-ray activity.  The stellar
systems listed have $X$-ray spectra which are neither periodic
(that would correspond to a rotating neutron star), nor recurrent
(which is interpreted as thermonuclear explosions on a neutron
star's hard surface). The final selection criterion is that of the
mass $M_c$ exceeding the Chandrasekhar limit $M_C\approx 3$ solar
masses $ M_{\odot}$.\footnote{See~\cite{Narayan} for a discussion
and
  references concerning the value of $M_C$.} According to the authors
of~\cite{CMSciama}, the strongest black hole candidate in 1999 was
V404 Cygni, which belongs to the LMXB class.

\begin{table}
 \caption{{Stellar mass black hole candidates (from
\cite{Luminet})}}
\begin{center}
  \renewcommand{\arraystretch}{1.4} \setlength\tabcolsep{5pt}
\begin{tabular}{llll}
 \hline\noalign{\smallskip}
 Type & Binary system  &  $M_{c}/M_{\odot}$ & $M_{*}/M_{\odot}$     \\
   \noalign{\smallskip} \hline \noalign{\smallskip}
 {HMXB:}    &  Cygnus X-1    &  11--21 &  24--42  \\
&LMC X-3   & 5.6 --7.8 &  20  \\
&LMC X-1      & $\geq$ 4 &  4--8 \\
   \noalign{\smallskip} \hline \noalign{\smallskip}
 {LMXB: }       &            V 404 Cyg        & 10--15 & $\approx$ 0.6
 \\ &A 0620-00        &  5--17 &  0.2--0.7 \\
&GS 1124-68 (Nova Musc)      &  4.2--6.5 &  0.5--0.8  \\
&GS 2000+25 (Nova Vul 88)   &  6-14 & $\approx$ 0.7  \\
&   GRO J 1655-40  & 4.5 -- 6.5 & $\approx$ 1.2\\
&  H 1705-25 (Nova Oph 77)     &  5--9 &  $\approx$ 0.4\\
&J 04224+32      & 6--14 & $\approx$ 0.3 -- 0.6  \\
\hline
\end{tabular}
\end{center}\label{luminet:B}
\end{table}

Table~\ref{luminet:B} should be put into perspective by realizing
that, by some estimates~\cite{Luminet}, a typical galaxy -- such
as ours -- should harbour $10^7-10^8$ stellar black mass holes.
We note an interesting proposal, put forward in~\cite{CGME}, to
carry out observations by gravitational microlensing of some 20
000 stellar mass black holes that are predicted~\cite{MG} to
cluster within 0.7 pc of Sgr A$^*$ (the centre of our galaxy).
 \begin{table}
\renewcommand{\arraystretch}{1.4}
\setlength\tabcolsep{5pt} \caption{{Twenty-nine supermassive black
hole candidates (from \cite{MerFer,KorGeb})}}
\begin{center}
\begin{tabular}{lllll}
\hline\noalign{\smallskip} dynamics of     & host galaxy  & $M_{h}/M_{\odot}$  & host galaxy  & $M_{h}/M_{\odot}$   \\
\noalign{\smallskip} \hline \noalign{\smallskip}
water maser discs:    
   & NGC 4258             & $ 4 \times 10^{7}$ && \\
\noalign{\smallskip} \hline \noalign{\smallskip}
gas discs:   & IC 1459 &  $2 \times 10^{8}$ & M 87             & $ 3 \times 10^{9}$ \\
   & NGC 2787           & $ 4 \times 10^{7}$ & NGC 3245             & $ 2 \times 10^{8}$ \\
       & NGC 4261         & $ 5 \times 10^{8}$ & NGC 4374         & $ 4 \times 10^{8}$\\
       & NGC 5128         & $ 2 \times 10^{8}$ & NGC 6251         & $ 6 \times 10^{8}$\\
              & NGC 7052       & $ 3\times 10^{8}$ & &\\
\noalign{\smallskip} \hline \noalign{\smallskip}
stars:     
     & NGC 821         & $ 4 \times 10^{7}$& NGC 1023         & $ 4 \times 10^{7}$\\
     & NGC 2778          & $ 1 \times 10^{7}$& NGC 3115          & $ 1 \times 10^{9}$\\
     & NGC 3377         & $ 1 \times 10^{8}$
     & NGC 3379         & $ 1\times 10^{8}$\\
     & NGC 3384        & $ 1 \times 10^{7}$
     & NGC 3608         & $ 1\times 10^{8}$\\
     & NGC 4291         & $ 2 \times 10^{8}$
     & NGC 4342        & $ 3\times 10^{8}$\\
     & NGC 4473         & $ 1 \times 10^{8}$
     & NGC 4486B          & $ 5 \times 10^{8}$\\
     & NGC 4564         & $ 6 \times 10^{7}$
     & NGC 4649         & $ 2 \times 10^{9}$\\
     & NGC 4697        & $ 2 \times 10^{8}$
     & NGC 4742          & $ 1 \times 10^{7}$\\
     & NGC 5845         & $ 3 \times 10^{8}$
     & NGC 7457    & $ 4 \times 10^{6}$\\
     & Milky Way              & $ 2.5 \times 10^{6}$\\
\noalign{\smallskip}\hline
\end{tabular}
\end{center}\label{luminet:BB}
\bhindex{gravitational!supermassive}
\end{table}

\begin{figure}[b]
\begin{center}
  \includegraphics[width=.7\textwidth]{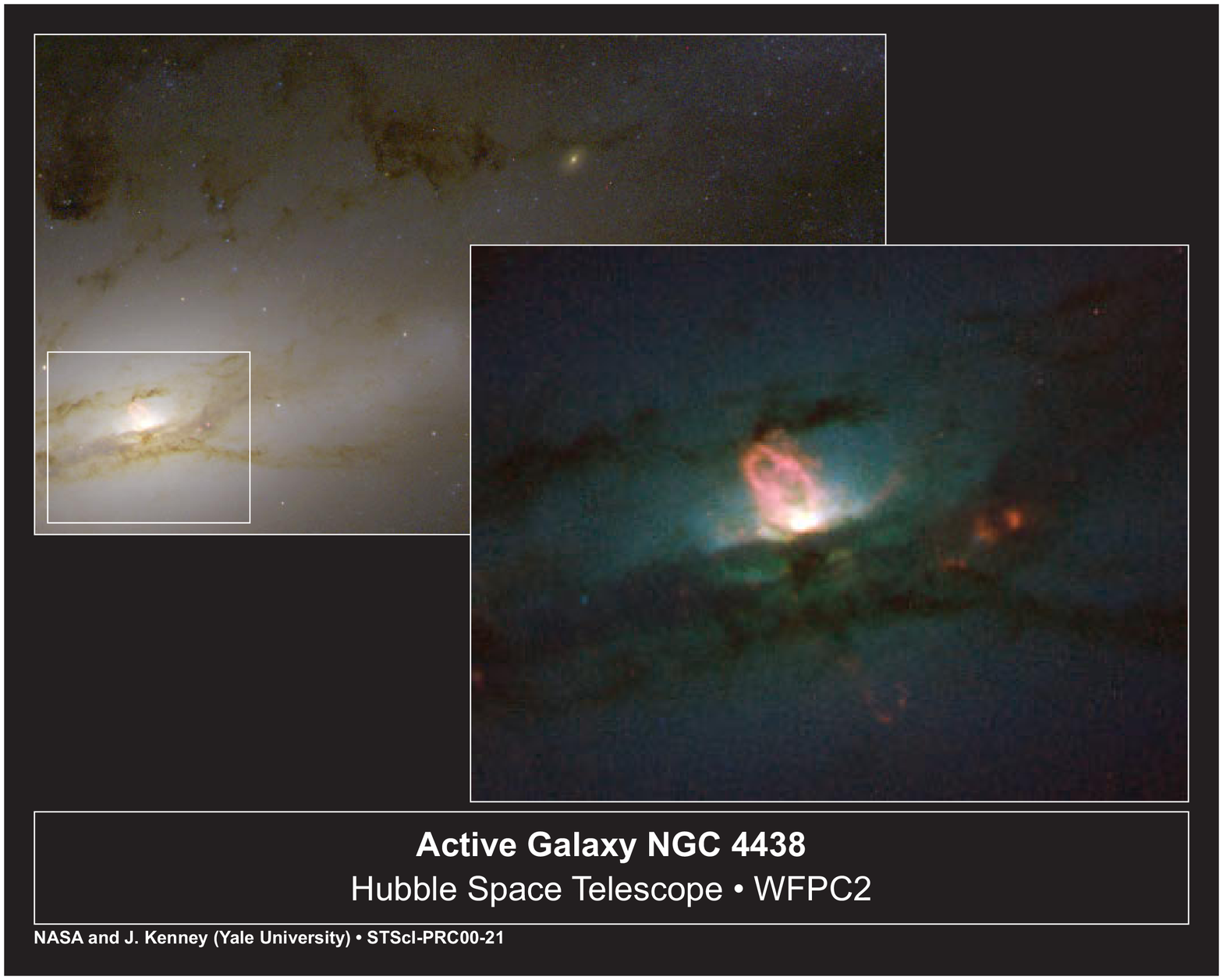}
\end{center}
\caption{{Hubble Space Telescope observations~\cite{Kenney} of the
    nucleus of the galaxy NGC 4438, from the STScI Public
    Archive~\cite{STScI}.}} \label{ngc4438}
\end{figure}

It is now widely accepted that quasars and active galactic nuclei
are powered by accretion onto massive black holes
\cite{deZeeuw,Madejski}. Further, over the last few years there
has been increasing evidence that massive dark objects may reside
at the centres of most, if not all, galaxies
\cite{Macchetto,ReesSupermassive}. In several cases the best
explanation for the nature of those objects is that they are
``heavyweight" black holes, with masses ranging from $10^6$ to
$10^{10}$ solar masses. Table~\ref{luminet:BB}\footnote{The table
  lists those galaxies which are listed both in~\cite{MerFer}
  and~\cite{KorGeb}; we note that some candidates from earlier
  lists~\cite{Rees:Wald} do not occur any more
  in~\cite{MerFer,KorGeb}. Nineteen of the observations listed have
  been published in 2000 or 2001.} lists some supermassive black hole
candidates; some other candidates, as well as precise references,
can be found in~\cite{Narayan,MerFer,KorGeb,Rees:Wald}. The main
criterion for finding candidates for such black holes is the
presence of a large mass within a small region; this is determined
by maser line spectroscopy, gas spectroscopy, or by measuring the
motion of stars orbiting around the galactic nucleus. The reader
is referred to~\cite{MGH} for a discussion of the maser emission
lines and their analysis for the supermassive black hole candidate
NGC 4258.  An example of measurements via gas spectrography is
given by the analysis of the
\begin{figure}[b]
\begin{center}
  \includegraphics[width=.5\textwidth]{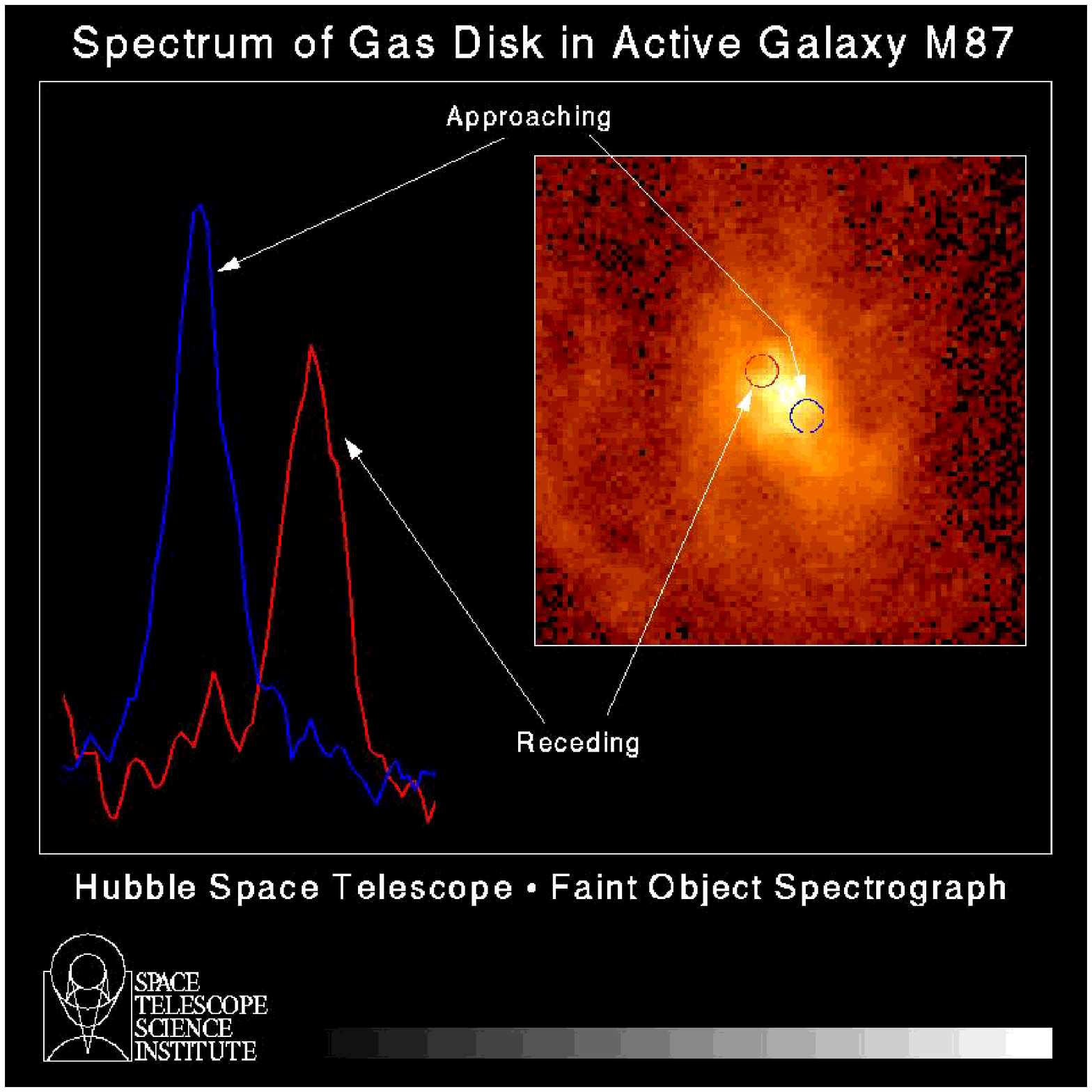}
\caption{Hubble Space Telescope observations of spectra of gas in
  the vicinity of the nucleus of the radio galaxy M~87, NASA and H.
  Ford (STScI/JHU)~\cite{STScI}.}\label{m87}
\end{center}\bhindex{gravitational!spectrographic observations of}
\end{figure}%
Hubble Space Telescope (HST) observations of the radio galaxy
M~87~\cite{harms} (compare~\cite{Madejski}): A spectral analysis
shows the presence of a disk-like structure of ionized gas in the
innermost few arc seconds in the vicinity of the nucleus of M~87.
The velocity of the gas measured by spectroscopy
(cf.~Fig.~\ref{m87}) at a distance from the nucleus of the order
of $ 6 \times 10^{17}$~m, shows that the gas recedes from us on
one side, and approaches us on the other, with a velocity
difference of about $ 920$~km~s$^{-1}$ . This leads to a mass of
the central object of $\sim 3 \times 10^{9}$~$M_{\odot}$, and no
form of matter can occupy such a small region except for a black
hole. Figure~\ref{ngc4438} shows another image, reconstructed out
of HST observations, of a recent candidate for a supermassive
black hole -- the (active) galactic nucleus of NGC 4438~\cite{Kenney}.%
To close the discussion of Table~\ref{luminet:BB}, we note that
the determination of mass of the galactic nuclei via direct
measurements of star motions has been made possible by the
unprecedentedly high angular resolution and sensitivity of the
HST. There seems to be consensus
\cite{ReesSupermassive,KorGeb,MerFer} that the two most convincing
supermassive black hole candidates are the galactic nuclei of
NGC~4258 and of our own Milky Way.

There is a huge gap between the masses of stellar mass black holes
and those of the supermassive black holes and, in Bromley's
terminology~\cite{Bromley}, ``middleweight" candidates would be
welcome.  Such a tentative black hole with $M$ of the order of $
460M_{\odot}$ has been identified in M~82 by A.~Ptak and
R.~Griffiths~\cite{PtakGriffiths}. E.~Colbert and
R.~Mushotzky~\cite{ColbertMushotzky} give a list of compact
$X$-ray sources which could include some further black holes with
masses in the $10^2-10^4 M_{\odot}$ range.

Let us close this section by pointing out the review
paper~\cite{Carr} which discusses both theoretical and
experimental issues concerning \emph{primordial} black holes.

\bhindex{gravitational!astrophysical|)}

\section{Causality for symmetric hyperbolic systems}\label{Scshs}
\chnindex{causality|(} \chnindex{symmetric hyperbolic systems} The
usual, Scri based, definition of a black hole proceeds as follows:
one starts by introducing a preferred asymptotic region -- this is
usually taken to be a conformal boundary. Then the black hole
region is defined as the set of points which cannot send
information to the asymptotic region; we will return to the
details of this construction in Section~\ref{Ssbh} below. The key
ingredient here is the notion of ``not being able to send
information" -- this is usually defined using the metric, together
with the postulate that the propagation speed for objects carrying
information is bounded from above by that of null geodesics. The
aim of this section is to provide a precise mathematical version
of this notion, via Proposition~\ref{Puni} below, for any
symmetric hyperbolic system of differential equations, without
introducing any supplementary postulates. The point is that every
physical theory defines its own causality via the corresponding
system of equations. We will then show in Section~\ref{Sdbh} how
the ``Unruh metric" arises in the application of our construction
to the Euler equations.

Let, thus, $L$ be a quasi-linear, symmetric hyperbolic, first
order partial differential operator acting on sections $\varphi$
of a bundle $E$, with scalar product $\langle\cdot,\cdot\rangle$,
over a manifold $\mcM$; in a local trivialization over a
coordinate patch $\mcU\ni x^\mu$ we have \bel{sh1} L[\varphi]:=
A^\mu(\varphi,x^\beta)\partial_\mu \varphi + B(\varphi)\;,\ee
where the $A^\mu$'s are endomorphisms of the fibers of $E$,
symmetric with respect to $\langle\cdot,\cdot\rangle$. Given a
section $\varphi$ of $E$ (defined perhaps on a subset of $\mcM$),
a hypersurface $\hyp$ is said to be {\em spacelike}\footnote{In
quasi-linear theories, in which the coefficients of the highest
derivatives depend upon the fields, the notion of spacelikeness
will of course depend upon the given set of fields.} if one of its
fields of co-normals $n_\mu$ satisfies
$$A^\mu\left(\varphi\left(p\right),p\right)n_\mu(p) >0\;,\quad
p\in\hyp\;;$$ this is understood in the sense of strict positive
definiteness of endomorphisms. By definition of a symmetric
hyperbolic system, every point of the domain of definition of
$\varphi$ lies on some spacelike hypersurface. For $p\in\mcM$ we
set
\begin{subeqnarray}
  \timkd_p & := & \{\alpha\in T^*_p\mcM \;|\;
  A^\mu\left(\varphi\left(p\right),p\right)\alpha_\mu \ \textrm{ is
    positive definite} \}\subset T^*_p\mcM
  \; , \label{f1a}\\
  \cimkd_p &:=&\overline{\timkd_p} \subset \{\alpha\in T^*_p\mcM \;|\;
  A^\mu\left(\varphi\left(p\right),p\right)\alpha_\mu \ \textrm{ is
    non-negative definite} \}\subset T^*_p\mcM \; , \nonumber \\
  &&\label{f1b}
\end{subeqnarray}
and \bel{f2}\timkd := \cup_{p\in\mcM}\timkd_p\subset T^*\mcM\;,
\qquad \cimkd := \cup_{p\in\mcM}\cimkd_p=\overline{\timkd}\;.\ee
The bundle $\timkd$ is thus the bundle of covectors normal to
spacelike hypersurfaces. Because positive definite matrices form a
convex set, each set $\timkd_p$ is a convex cone in $T^*_p\mcM$.
The bundles $\timku\subset T\mcM$, respectively $\cimku\subset
T\mcM$, of \emph{timelike}, respectively \emph{causal}, vectors
are defined by duality:
\begin{subeqnarray}
  \timku_p & := & \{X\in T_p\mcM \;|\forall\; \alpha \in \timkd_p\
  X^\mu\alpha_\mu >0\}
  \; , \label{f1c}\\
  \cimku_p &:=&\overline{\timku_p} = \{X\in T_p\mcM \;|\forall\;
  \alpha \in \timkd_p\ X^\mu\alpha_\mu \ge 0\} \; . \label{f1d}
\end{subeqnarray}
Recall that all standard classical field equations, i.e., Euler's
equations, the scalar wave equation, the wave map equation,
Maxwell's equations, Yang-Mills equations, or Einstein's
equations, can be written as a symmetric hyperbolic system ({cf.,
e.g.},~\cite{Racke} or \cite[Vol.~IV]{Taylor}). For example, the
wave equation for a scalar field $u$ on a Lorentzian manifold
$(\mcM,g)$ can be written in the form \eq{sh1} by introducing a
new set of variables
$$\varphi:=(u,e_\mu(u))\;,$$
where the $e_\mu$'s, $\mu=0,\ldots,n$, form an ON-frame for $g$.
In this case $\timku$ coincides with the bundle of timelike future
pointing vectors of the metric $g$ (where the future direction is
determined by that of $e_0$), $\cimku$ coincides with the bundle
of causal future pointing or vanishing vectors of $g$, while
$\timkd$ and $\cimkd$ are, modulo sign (we are using the signature
$(-,+,\ldots,+)$), ``$\timku$ and $\cimku$ with indices lowered
using the metric". The same remains true for wave maps, for linear
electrodynamics (considered as a first order system for $\vec E$
and $\vec B$), for the Yang-Mills equations in the Lorentz gauge,
and for the usual symmetric hyperbolic reduction of Einstein's
equations based on harmonic coordinates. (See the end of this
section for more comments about the Einstein case.)

We emphasize that the bundles $\timku$, etc., are defined solely
by the system of equations under consideration, independently of
any metric. They can be used to determine the causality properties
of a given symmetric hyperbolic system by mimicking the usual
metric constructions
\cite{PenroseDiffTopo,Beem-Ehrlich:Lorentz,HE,ONeill}:
\emph{timelike future directed paths} are defined as piecewise
differentiable maps $\gamma$ from $ I$ to $\mcM$, where $I\subset
\R$ is an interval, with the property that the tangent $\dot
\gamma$ is in $\timku$ wherever defined; \emph{causal future
directed paths} are defined as Lipschitz continuous paths such
that $\dot \gamma$ is in $\cimku$ wherever defined. \emph{Past
directed paths} are paths which are obtained from future directed
paths by a reversal of parameterization. The \emph{timelike
future} $I^+(\Omega)$, respectively the \emph{causal future}
$J^+(\Omega)$, of a set $\Omega\subset \mcM$ is defined as the set
of points which can be reached form $\Omega$ by following a future
directed timelike, respectively causal, path. Causal and timelike
futures $J^-$ and $I^-$ are obtained form the above definition by
replacing "future" by "past". Notions such as global
hyperbolicity, and so on, can be defined in the usual way. It
might be convenient to write $I^\pm(\Omega,L)$, etc., to emphasize
the dependence upon the system of equations $L$, if ambiguities
can arise. We note that for semi-linear systems -- by definition,
these are the systems in which the coefficients $A^\mu$ in
\eq{sh1} do not depend upon $\varphi$ -- the resulting causal
constructs are independent of the solution $\varphi$ under
consideration; however, this will not be the case in general.

We are ready now to address the issue of main interest in this
work -- the notion of a black hole. Given a set $\Omega\subset
\mcM$ and a symmetric hyperbolic system $L$ we define the
\emph{black hole region
  $\mcB_\Omega(L)$ of $\Omega$} as \bel{bhr} \mcB_\Omega(L):=
\mcM\setminus J^-(\Omega,L)\;.\ee The object so defined depends
upon the set $\Omega$, and acquires its full meaning when $\Omega$
is naturally distinguished by the problem at hand.

In several cases, some of which have already been mentioned, the
causality theory constructed above arises out of a metric; we will
show in Section~\ref{Sdbh} below that this holds also for the
Euler equations. The intuitive meaning associated to the notion of
a black hole region is that no information from $\mcB_\Omega(L)$
can reach $\Omega$. This is made precise by the following result,
the proof of which proceeds by causality arguments which are known
in principle:

\begin{ptcproposition}\label{Puni} Let $\varphi$ be a solution of
  \bel{Puni1} L[\varphi]=J\;,\ee where $L$ is symmetric hyperbolic
  operator \eq{sh1} on a manifold $\mcM$, and suppose that there
  exists a Lorentzian metric $g$ such that the sets $\cimku$ defined
  in \eq{f1d} are future light cones of $g$. Assume that $(\mcM,g)$ is
  globally hyperbolic with Cauchy surface $\hyp$ and consider a set
  $\Omega\subset\mcM$ such that $\mcB_\Omega(L) \ne \emptyset$.  If
  $\varphi'$ is a solution of \bel{Puni2} L[\varphi']=J' \ \textrm{
    with }\ \varphi|_{\hyp}= \varphi'|_{\hyp}\;,\ee and if the
  difference of the source terms $J-J'$ is supported in
  $\mcB_\Omega(L)$, then $\varphi=\varphi'$ on $J^-(\Omega,L)$.
\end{ptcproposition}

We believe that the conclusion above is true \emph{without} the
assumption that causality is determined by a metric, but we are
not aware of any studies of this question in the current context;
it would be of interest to settle this. The reader is referred
to~\cite{Leray} for an analogous analysis of strictly hyperbolic
systems, and to~\cite{Christodoulou:action} for that of second
order Lagrangean hyperbolic systems.

We close this section by mentioning that care has to be taken when
interpreting the notions above for systems of equations which are
not directly in first order symmetric hyperbolic form: it can
happen that the physical equations can be then rewritten in
several different symmetric hyperbolic forms, leading to different
notions of pasts, futures, black hole regions, etc. This does
indeed happen for the Einstein equations: in
\cite[Section~4.1]{Friedrich:2000qv} a symmetric hyperbolic system
is presented which can be used to solve the Einstein equations and
in which the causal cones, as defined above, differ from the
metric ones; see the discussion after Equation~(4.25) there.
Another example of this kind arises if the $T$-vector of the
system of equations given in Sect.~5.2
of~\cite{Friedrich:hyperbolicreview} is chosen to be spacelike.
This is related to the acausal (in the usual metric sense)
propagation of gauge degrees of freedom in the associated systems,
and does of course not affect the propagation of any physically
relevant quantities. Given a system of equations, say $(S)$, a
simple way out of this problem is to define the physical black
hole region as the intersection, as $L$ runs over the set of those
symmetric hyperbolic systems $L$ the solutions of which include
solutions of $(S)$, of the associated black hole regions
$\mcB_\Omega(L)$. We note that the existence of exact solutions
such as the pp-waves, where nonlinear perturbations propagate
along null geodesics, guarantees that every symmetric hyperbolic
system $L$ which reproduces the Einstein's equations will
necessary contain the metric light cone as its causal light cone
associated to $L$. It follows that the above prescription does
indeed reproduce the usual metric definition of a black hole
region for the vacuum Einstein equations.

\subsection{Dumb holes}\label{Sdbh}
\chindex{dumb holes} \bhindex{sonic} Let us illustrate the
considerations of the previous section in the case of the Euler
equations \bea\nonumber & \rho \left(\dot {\bf v} +{\bf v}\cdot
\nabla
  {\bf v} \right)+\nabla p(\rho) = 0\;,& \\& \dot\rho +
\nabla(\rho{\bf v})=0\;.& \label{Euler}\eea For strictly positive
$\rho$'s  \eq{Euler} can be rewritten as a symmetric hyperbolic
system of the form \eq{sh1} with $B=0$, for a field
$\varphi=(v^i,\rho)$, by introducing the matrices \bel{ptc:E1}
A^0=\left(\begin{array}{cc} \rho\id_{\R^3} & 0
    \\ 0 &\frac{p'}{\rho} \end{array}\right)\;,\quad
A^i=\left(\begin{array}{cc} \rho v^i\id_{\R^3}& p'\delta^i_j \\
    p'\delta^i_j &\frac{p'}{\rho}v^i \end{array}\right)\;, \ee which
are clearly symmetric with respect to the standard scalar product
on $\R^3\times\R\ni (\vec v,\rho)$. Straightforward algebra shows
that $A^\mu n_\mu$ is strictly positive if and only if $\rho>0$,
$dp/d\rho>0$ and \bel{Eul3} -(n_0+v^in_i)^2+ p' \sum_i (n_i)^2 < 0
\quad \Longleftrightarrow \quad g^{\mu\nu}n_\mu n_\nu < 0\;,\ee
where the {\em Unruh metric} $g$ is defined
as~\cite{Unruh,Visser:1998qn}
%
%
%
%
\bel{Eu4} g^{\mu\nu}= \frac1{c_s\rho} \left(\matrix{-1  & -v^i\cr
-v^j & c_s^2\delta^{ij}-v^iv^j\cr}\right)\ \Longleftrightarrow\
g_{\mu\nu}= \frac\rho {c_s}\left(\matrix{-c_s^2+{\vec v^2} &
-v^i\cr -v^j & \delta_{ij}\cr}\right)\;, \ee
%
%
%
%
\[
c_s^2 = {dp\over d\rho}\;.
\]
This metric exhibits a typical black hole behaviour for stationary
solutions in which the speed $\vec v$ of the fluid  velocity meets
the speed of sound $c_s$ across a smooth hypersurface.

The original argument of Unruh leading to \eq{Eu4} was a
perturbational one, for irrotational flows:
\begin{equation}
  \vec{v} = \nabla\Phi\;.
\end{equation}
Small perturbations $\phi\equiv \delta \Phi$ of $\Phi$ satisfy
then a scalar wave equation~\cite{Unruh},
\[
\partial_\mu (\sqrt{-g} g^{\mu\nu}\partial_\nu \phi) = 0\; ,
\]
in the metric \eq{Eu4}. Our discussion above shows that neither
the irrotationality of the flow, nor the perturbative character of
the argument are essential for the problem at hand. The fact that
causality for the full non-linear Euler's equations is governed by
a metric can essentially be found in Courant and Hilbert, and is
certainly well known to some researchers (H.~Friedrich, private
communication).

There was renewed interest in the above because of the recent
experiments with the Bose-Einstein condensates. Recall that in the
Madelung formulation, the Gross-Pitaevskii equation governing the
dynamics of the condensates can be rewritten in  Euler form. There
have been suggestions that some of the effects associated to
black-hole type causality, including the analogue of the Hawking
radiation, could be observed in such systems, see
\cite{Volovik:2000ua,Jacobson:1999zk,Barcelo:2001ca} and
references therein. While the resulting black holes are still
referred to as \emph{sonic}, this is rather misleading as the
underlying Gross-Pitaevskii equation does not have anything to do
with sound propagation in liquid or gaseous physical media.

\subsection{Optical holes}
\chindex{optical holes} \bhindex{optical} According to
M.~Visser~\cite{Visser:mog}, in a dielectric fluid ``the
refractive index $n$, the fluid velocity $v$, and the background
Minkowski metric $\eta$ can be combined algebraically to provide
an effective metric $g$''
$$
g_{\mu\nu} = \eta_{\mu\nu} - (n^2-1)v_\mu v_\nu \;. $$ This leads
to black-hole effective geometries for refractive indices which
exceed 1 in some regions. Phenomenological models for this
behaviour have been proposed by U.~Leonhardt and P.~Piwnicki,
cf.~\cite{Leonhardt:Piwnicki} and references therein\footnote{See
also
  URL~\url{http://www.st-and.ac.uk/~www_pa/group/quantumoptics}}.
Similarly, a (different) effective metric is obtained in nonlinear
electrodynamics, leading -- according to~\cite{Novello:2001fv} --
to models which sometimes contain closed timelike curves. The
status of these ``analogous models" seems to be somewhat less
clear than that of the sonic ones.

\subsection{Trapped surfaces}
\chindex{trapped surface} We shall close this section by noting
the discussion in~\cite{Leonhardt:2000hf,Visser:2000pk} concerning
the notion of trapped surfaces\footnote{Cf. Section~\ref{Sah} for
the
  differential-geometric definition of this notion} for black hole
geometries -- such surfaces are useful detectors of black hole
regions in GR with matter fields carying positive energy. It
should be borne in mind that while causality concepts can be
introduced for any symmetric hyperbolic system, the existence of a
trapped surface signals the presence of a black hole region
\emph{only} when causality is governed by \emph{a metric
satisfying an energy positivity
  condition} (and, if in a Scri context, when Scri satisfies various
regularity conditions, see Section~\ref{Ssbh} below). In the
non-gravitational black hole models the metric is the one arising
out of the causality structure of the theory, and there are no
reasons in general to believe that its Ricci tensor should have
any preferred properties; to start with, it is defined only up to
a conformal factor anyway. In addition, the "trapped surface"
terminology is used in~\cite{Visser:2000pk} for objects which are
completely unrelated to the usual differential geometric context,
which is extremely misleading and obscures the problems at hand.
\chnindex{causality|)}

\section{Standard black holes}\label{Ssbh}
\bhindex{gravitational!Scri based|(} The standard way of defining
black holes is by using conformal completions: A pair $(\tcM
,\cg)$ is called a \emph{conformal completion} of $(\mcM,g)$ if
$\tcM $ is a manifold with boundary such that:
\begin{enumerate}\item $\mcM$ is the interior of $\tcM $, \item there exists a function $\Omega$, with the
property that the metric $\tg$, defined to be $\Omega^2g$ on
$\mcM$, extends by continuity to the boundary of $\tcM$, with the
extended metric still being non-degenerate throughout, \item
$\Omega$ is positive on $\mcM$, differentiable on $\tcM $,
vanishes on $\Scri$, with $d\Omega$ \emph{nowhere vanishing} on
$\Scri$.
\end{enumerate} We emphasize that no assumptions about the causal
nature of Scri are made so far. The boundary of~$\tcM $ will be
called Scri, denoted $\Scri$. In the usual
textbooks~\cite{HE,Wald:book} smoothness of both the conformal
completion and the metric $\cg$ is imposed, though this can be
weakened for many purposes. We set
$$
\Scri^+=\{p\in\Scri\ |\ I^{-}(p;\tcM )\cap \mcM \ne \emptyset\}\ .
$$
Assuming various global regularity conditions on $\tcM $, to which
we shall return in \Sect{Scrc}, one then defines the black hole
region $\mcB$ as \bel{sbh1} \mcB:= \cM\setminus J^-(\scrip)\;.\ee
Let us point out some drawbacks of this approach:

\begin{itemize}
\item {\bf Non-equivalent Scri's:} \chindex{Scri!non-equivalent}
  Conformal completions at null infinity do not have to be unique, an
  example can be constructed as follows: the Taub--NUT metrics can be
  locally written in the form \cite{Misner}
\begin{eqnarray}
& -U^{-1}dt^2  +(2L)^2U\sigma^2_1 + (t^2 + L^2)(\sigma^2_2 +
\sigma^2_3)\,, \label{MBianchi} & \\ & U(t) = -1 + {2(mt +
L^2)\over
  t^2 + L^2}\,. \label{UBianchi}
\end{eqnarray}
where $\sigma_1$, $\sigma_2$ and $\sigma_3$ are left invariant
one--forms on $SU(2)\approx S^3$.  The constants $L$ and $m$ are
real numbers with $L > 0$. Parameterizing $S^3$ with Euler angles
$(\mu,\theta,\varphi)$ one is led to the following form of the
metric
 \begin{eqnarray*}
& g= -U^{-1}dt^2 +(2L)^2U(d\mu + \cos\theta d\varphi)^2 + (t^2 +
L^2)(d\theta^2+\sin^2\theta d\varphi^2)\,. 
\end{eqnarray*}
The standard way of performing extensions across the Cauchy
horizons $t_\pm:= M \pm \sqrt{M^2+L^2}$ is to introduce new
coordinates \bel{UB4}(t,\mu,\theta,\varphi)\to(t,\mu\pm
\int_{t_0}^t [2L U(s)]^{-1} ds, \theta,\varphi)\;,\ee which gives
\begin{eqnarray} g_{\pm}
&=& \pm 4L(d\mu + \cos\theta d\varphi)dt \nonumber \\
&&+(2L)^2U(d\mu + \cos\theta d\varphi)^2+ (t^2 +
L^2)(d\theta^2+\sin^2\theta d\varphi^2)\,. \label{UB3}
\end{eqnarray}
Each of the metrics $g_{\pm}$ can be smoothly conformally extended
to the boundary at infinity ``$t=\infty$" by introducing
$$x=1/t\;,$$ so that \eq{UB3} becomes
\begin{eqnarray} g_{\pm}
&=& x^{-2}\Big(\mp 4L(d\mu + \cos\theta d\varphi)dx \nonumber \\
&&+(2L)^2x^2U(d\mu + \cos\theta d\varphi)^2+ (1 +
L^2x^2)(d\theta^2+\sin^2\theta d\varphi^2)\Big)\,. \label{UB5}
\end{eqnarray}
In each case this leads to a Scri diffeomorphic to $S^3$. There is
a simple isometry between $g_+$ and $g_-$ given by
$$
(x,\mu,\theta,\varphi)\to(x,-\mu,\theta,-\varphi)$$ (this does
correspond to a smooth map of the region $t\in(t_+,\infty)$ into
itself, cf.~\cite{ChImaxTaubNUT}), so that the two Scri's so
obtained are isometric. However, in addition to the two ways of
attaching Scri to the region $t\in(t_+,\infty)$ there are the two
corresponding ways of extending this region across the Cauchy
horizon $t=t_+$, leading to four possible manifolds with boundary.
It can then be seen, using e.g.~the arguments
of~\cite{ChImaxTaubNUT}, that the four possible manifolds split
into two pairs, each of the manifolds from one pair \emph{not}
being isometric to one from the other. Taking into account the
corresponding completion at ``$t=-\infty$", and the two extensions
across the Cauchy horizon $t=t_-$, one is led to four inequivalent
conformal completions of each of the two
inequivalent~\cite{ChImaxTaubNUT} time-oriented, maximally
extended, standard Taub-NUT space-times.

\hspace{.4cm} Under mild completeness conditions in the spirit of
\cite{GerochHorowitz}, uniqueness of $\scrip$ as a point set in
{\em
  past-distinguishing} space-times should follow from the TIP and TIF
construction of~\cite{GKP}; however, uniqueness of Scri's
differentiable structure within that framework is far from being
clear.

\hspace{.4cm} Yet another approach to uniqueness has been proposed
by Geroch~\cite{GerochEspositoWitten}: A completion as defined at
the beginning of this section is said to be \emph{a regular
asymptote} if, given any point $p\in\scri$, and any non-zero null
vector $\ell\in T_q\tcM $ such that the maximally extended null
geodesic with tangent $\ell$ does not meet p, then there are
neighbourhoods $\mcU\subset\tcM $ of $p$ and $\mcV\subset T\tcM $
of $\ell$ such that no maximally extended null geodesic with
tangent in $\mcV$ meets $\mcU$. This regularity condition is not
satisfied by the Taub-NUT completions described above. Supposing
that the set of regular asymptotes is non-empty,
Geroch~\cite[Theorem~2, p.~14]{GerochEspositoWitten} gives an
argument for existence of a maximal regular asymptote, unique up
to well behaved conformal transformations. Now, we have not been
able to fill in the details\footnote{It is not completely clear
that
  the map $\varphi$ defined by Geroch in his proof is simultaneously
  differentiable for all regular asymptotes, as asserted
  in~\cite[p.~14]{GerochEspositoWitten}. The regularity condition
  guarantees Hausdorffness of the maximal regular asymptote
  constructed in~\cite[p.~14]{GerochEspositoWitten}, but does not seem
  to guarantee its differentiability in any obvious way.} of some of
the arguments suggested in~\cite{GerochEspositoWitten}, but the
construction given is plausible enough so that we expect that it
might actually be correct. Whatever the case, Geroch's regularity
requirement is a global condition which seems to be difficult to
control in general situations.\footnote{Globally hyperbolic
conformal
  completions (in the sense of manifolds with boundary) should satisfy
  Geroch's condition. Nevertheless, it should be borne in mind that
  good causal properties of a space-time might fail to survive the
  process of adding a conformal boundary. For example, adding a Cauchy
  horizon to a maximal globally hyperbolic space-time sometimes leads
  to space-times which are not globally hyperbolic in the sense of
  manifolds with boundary.} In particular it could happen that many
space-times of interest admitting conformal completions do not
admit any regular ones. It should be borne in mind that one of the
main applications of the $\scri$ framework is the possibility of
\emph{uniquely} defining global charges such as mass, angular
momentum, etc. (cf.~\cite{CJK} for a new approach to this
question), in which issues about global behaviour of \emph{all}
geodesics seem completely irrelevant. The fact that there is no
suitable uniqueness result for the differentiable structure of
conformal completions is a serious gap in our understanding of
objects such as the Trautman-Bondi mass.

\hspace{.4cm} We note that uniqueness of a class of Riemannian
conformal completions at infinity has been established
in~\cite[Section~6]{ChHerzlich}; this result can probably be used
to obtain uniqueness of differentiable structure of Lorentzian
conformal completions for Scri's admitting  cross-sections.
Further partial results on the problem at hand can be found
in~\cite{Schmidtcqg91}.

\item {\bf Poorly differentiable Scri's:}
  \chindex{Scri!differentiability of} In all standard
  treatments~\cite{HE,Wald:book,GerochEspositoWitten} it is assumed
  that both the conformal completion $\tcM =\cM\cup\scri$ and the
  extended metric $\cg$ are smooth, or have a high degree of
  differentiability~\cite{penrose:scri}. This is a restriction which
  excludes most space-times which are asymptotically Minkowskian in
  lightlike directions, see~\cite{PRLetter,Kroon:detect} and
  references therein. Poor differentiability properties of $\scri$
  change the peeling properties of the gravitational
  field~\cite{ChMS}, but most -- if not all -- essential properties of
  black holes should be unaffected by conformal completions with,
  e.g., polyhomogeneous differentiability properties as considered
  in~\cite{ChMS,AndChDiss}. It should, however, be borne in mind that
  the hypothesis of smoothness has been done in the standard
  treatments, so that in a complete theory the validity of various
  claims should be reexamined. Some new results concerning existence
  of space-times with a poorly differentiable Scri can be found
  in~\cite{Lengard}.
\item {\bf The structure of $i^+$:} The current theory of black holes
  is entirely based on intuitions originating in the Kerr and
  Schwarzschild geometries. In those space-times we have a family of
  preferred ``stationary" observers which follow the orbits of the
  Killing vector field $\partial_t$ in the asymptotic region, and
  their past coincides with that of $\scrip$. It is customary to
  denote by $i^+$ the set consisting of the points $t=\infty$, where
  $t$ is the Killing time parameter for those observers. Now, the
  usual conformal diagrams for those space-times~\cite{MTW,HE} leave
  the highly misleading impression that $i^+$ is a regular point in
  the conformally rescaled manifold, which, to the best of our
  knowledge, is not the case. In dynamical cases the situation is
  likely to become worse. For example, one can imagine black hole
  space-times with a conformal diagram which, to the future of a
  Cauchy hypersurface $t=0$, looks as in Fig.~\ref{FigureRT.3}.
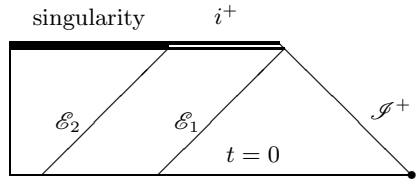
\begin{figure}
\begin{center}
\begin{picture}(-50,90)(100,30)
\thicklines
\put(-2,100){\line(1,0){102}} \put(-2,99){\line(1,0){60}}
\put(-2,98){\line(1,0){104}} \thinlines
\put(10,50){\line(1,1){48}} \put(-2,50){\line(1,0){152}}
\put(54,50){\line(1,1){48}} \put(-2,100){\line(0,-1){50}}
\put(150,50){\line(-1,1){50}}
\put(150,50){\circle*{3}} \put(90,55){\makebox(0,0)[b]{$t=0$}}
\put(134,75){\makebox(0,0)[l]{$\scrip$}}
\put(70,70){\makebox(0,0)[r]{$\mcE_1$}}
\put(25,70){\makebox(0,0)[r]{$\mcE_2$}}
\put(28,105){\makebox(0,0)[b]{singularity}}
\put(80,107){\makebox(0,0)[b]{$i^+$}}
\end{picture}
\label{iplus}
\end{center}
\caption[FigureRT.3]{An asymptotically flat space-time with an
  unusual $i^+$.} \label{FigureRT.3}
\end{figure}
In that diagram the set $i^+$ should be thought of as the addition
to the space-time manifold $\mcM$ of a set of points
``$\{t=\infty,q\in \mcO\}$", where $t\in[0,\infty)$ is the proper
time for a family of observers $\mcO$. The part of the boundary of
$\tcM $ corresponding to $i^+$ is a singularity of the conformally
rescaled metric, but we assume that it does \emph{not} correspond
to singular behaviour in the physical space-time. In this
space-time there is the usual event horizon $\mcE_1$ corresponding
to the boundary of the past of $\scrip$, which is completely
irrelevant for the family of observers $\mcO$, and an event
horizon $\mcE_2$ which is the boundary of the \emph{true} black
hole region for the family $\mcO$, i.e., the region that is not
accessible to observations for the family $\mcO$. Clearly the
usual black hole definition carries no physically interesting
information in such a setting.

\item {\bf Causal regularity of Scri:} As already pointed out, in
  order to be able to prove interesting results the definition
  \eq{sbh1} should be complemented by causal conditions on $\tcM $.
  The various approaches to this question,
  discussed in \Sect{Scrc}, are aesthetically highly unsatisfactory:
  it appears reasonable to impose causal regularity conditions on a
  space-time, but why should some unphysical completion have any such
  properties? Clearly, the physical properties of a black hole should
  not depend upon the causal regularity -- or lack thereof -- of some
  artificial boundary which is being attached to the space-time. While
  it seems reasonable and justified to restrict attention to
  space-times which possess good causal properties, it is not clear
  why the addition of artificial boundaries should preserve those
  properties, or even be consistent with them.  Physically motivated
  restrictions are relevant when dealing with physical objects, they
  are not when non-physical constructs are considered.
\item{\bf Inadequacy for numerical purposes:} Several\footnote{The
    numerical simulations
    in~\cite{Bishop:1997ik,Bartnik:NQSURL,Gomez:1998uj} cover regions
    extending all the way to infinity, within frameworks which seem to
    be closely related to the ``naive" framework of Section~\ref{Snbh}
    below.} numerical studies of black holes have been performed on
  numerical grids which cover finite space-time
regions~\cite{Alcubierre:2001vm,Alcubierre:2000ke,Kidder:2001tz,%
  Brandt:2000yp}. Clearly, it would be convenient to have a set-up
which is more compatible in spirit with such calculations than the
Scri one.
\end{itemize}
We will suggest, in Sections~\ref{Snbh} and \ref{Sqlbh} below, two
approaches in which the above listed problems are avoided. Before
doing this, let us complete the presentation of the usual, Scri
based, approach to black holes.
\subsection{Scri regularity conditions, and the area
  theorem}\label{Scrc} \chindex{Scri!regularity conditions|(} It is
easily seen that the definition~\eq{sbh1} is not very useful
without some {\em completeness conditions} on $\scrip$. For
instance, denote by $\tcM $ the standard conformal
completion\footnote{Inroduce
  coordinates $u=t-r$, $x=1/r$, $\theta$, $\varphi$ on
  $R^4\setminus\{r=0\}$, with $(r,\theta,\varphi)$ -- the usual
  spherical coordinates on $\R^3$, and set $\scrip:=\{x=0\}$ in those
  coordinates.} of the Minkowski space-time $\R^{1,3}$, with
$\scrip\approx \R\times S^2$, where the $\R$ factor is
parameterized by a Bondi coordinate $u$, with $u=t-r$ on
$\R^{1,3}$. One can obtain a new completion $\tcM _2$ by replacing
$\scrip$ by an open subset thereof, e.g.\ restricting the range of
$u$'s to $(-\infty,0)$. This will give a non-empty black hole
region $\mcB$ in $\R^{1,3}$,$$\mcB=J^+({0})\;,$$ which is clearly
a completely uninteresting statement. A way out of this problem
(as well as of some other related ones, discussed in
\cite{GerochHorowitz}) has been proposed\footnote{The ``WASE
setup"
  of~\cite{HE}, presented below, does not solve the problem,
  cf.~\cite{GerochHorowitz}.}  in~\cite{GerochHorowitz}, for
space-times which are vacuum near the conformal boundary, assuming
that the cosmological constant is zero: R.~Geroch and G.~Horowitz
introduce a preferred family of conformal factors $\Omega$ such
that the matrix $\Hess \Omega$ of second covariant derivatives of
$\Omega$ vanishes at $\scri$. It is then required that both
$\scrip$ and $\scri^-$ be diffeomorphic to $\R\times S^2$, with
the $\R$ factor corresponding to integral curves of
$\widetilde{\nabla }\Omega$, which are assumed to be complete.

Let us pass to a presentation of the causal regularity conditions
which are imposed in Hawking and Ellis's approach \cite{HE}. Now,
a set of conditions is only useful insofar as it allows to prove
something. Since the ``area theorem"\footnote{Recall that the area
  theorem is the statement that the area of cross-sections of event
  horizons is non-increasing towards the future. One possible precise
  version of this theorem can be found in Theorem~\ref{Tqlbharea}
  below.} is one of the landmark theorems in the theory of black
holes, we shall concentrate on the set of conditions from Hawking
and Ellis as used in their treatment of the area theorem; we
follow the presentation of~\cite[Appendix~B]{ChDGH}. One of the
conditions of the Hawking--Ellis area theorem
\cite[Proposition~9.2.7, p.~318]{HE} is that spacetime $({\mycal
M},g)$ is \emph{weakly asymptotically simple
  and empty} (``WASE'', \cite[p.~225]{HE}).  This means that there
exists an open set ${\mycal \mycalu}\subset \mycalm$ which is
isometric to $\mycalu'\cap \mycalm'$, where $\mycalu'$ is a
neighborhood of null infinity in an asymptotically simple and
empty (ASE) spacetime $(\mycalm',g')$ \cite[p.~222]{HE}.  It is
further assumed that $\mycalm$ admits a partial Cauchy surface
$\mycals$ with respect to which $\mycalm$ is \emph{future
asymptotically predictable} (\cite{HE}, p.~310). This is defined
by the requirement that $\scrip$ is contained in the closure of
the future domain of dependence ${\cal
  D}^+(\mycals; \mycalm)$ of $\mycals$, where the closure is taken in
the conformally completed manifold $\Mclosed =\mycalm \cup
\scrip\cup\scri^{-}$, with both $\scrip$ and $\scri^{-}$ being
null hypersurfaces. Next, one says that $(\mycalm,g)$ is
\emph{strongly
  future asymptotically predictable} (\cite{HE}, p.~313) if it is
future asymptotically predictable and if $J^{+}(\mycals)\cap \bar
J^{-}(\scrip;\Mclosed )$ is contained in ${\cal D}^+(\mycals;
\mycalm)$.  Finally (\cite{HE}, p.~318), $(\mycalm ,g)$ is said to
be a \emph{regular predictable space} if $(\mycalm ,g)$ is
strongly future asymptotically predictable and if the following
three conditions hold:
\begin{enumerate}
\item[($\alpha$)] $\mycals \cap \bar J^{-}(\scrip;\Mclosed )$ is
  homeomorphic to $\R^3\setminus$(an open set with compact closure).
\item [($\beta$)] $\mycals$ is simply connected.
\item [($\gamma$)] the family of hypersurfaces $\mycals(\tau)$
  constructed in \cite[Proposition~9.2.3, p.~313]{HE} has the property
  that for sufficiently large $\tau$ the sets $\mycals(\tau)\cap \bar
  J^{-}(\scrip;\Mclosed )$ are contained in $\bar
  J^{+}(\scri^{-};\Mclosed )$.
\end{enumerate}
It should be clear at this stage that this set of conditions is
rather intricate. As it turns out, it is not clear whether or not
it does suffice for a proof of the area theorem, as asserted in
\cite[Proposition~9.2.7, p.~318]{HE}: indeed, in the proof of
\cite[Proposition~9.2.1, p.~311]{HE} (which is one of the results
used in the proof of \cite[Proposition~9.2.7, p.~318]{HE}) Hawking
and Ellis write: ``This shows that if $\mycal{W}$ is any compact
set of $\mycals $, every generator of $\Scri^+$ leaves
$J^+({\mycal
  W};\Mclosed)$.'' The justification of this given in \cite{HE} is
wrong. If one is willing to impose the sentence in quotation marks
as \emph{yet one more regularity hypothesis} on $\scrip$, then the
arguments given in~\cite{HE} apply to prove the area theorem for
black holes with a piecewise smooth event horizon; the results
in~\cite{ChDGH} show that this remains true with no supplementary
conditions on the differentiability of the horizon. \chindex{area
  theorem}

Not only is the above set of hypotheses aesthetically unappealing,
it is far from being unique.  An alternative way to guarantee that
the area theorem will hold in the ``future asymptotically
predictable WASE'' set--up of \cite{HE} is to impose some mild
additional conditions on $\mycalu$ and $\mycals$. There are quite
a few possibilities, one such set of conditions has been described
in \cite[Appendix~B]{ChDGH}: Let $\psi : \mycalu \to \mycalu' \cap
\mycalm '$ denote the isometry arising in the definition of the
WASE spacetime $\mycalm $.  First, one requires that $\psi$ can be
extended by continuity to a continuous map, still denoted by
$\psi$, defined on $\overline \mycalu$.  Next, one demands that
there exists a compact set $K \subset \mycalm '$ such that,
\begin{equation}
  \label{condHE2}
  \psi(J^+(\mycals;\mycalm ) \cap \partial\mycalu) \subset J^+(K;\mycalm ')\;,
\end{equation}
see Fig.~\ref{condHE.fig}. \begin{figure}[t]
\begin{center}
  \input{HEcond.pstex_t}
\end{center}\caption[]{The set $\Omega\equiv
  J^+(\mycals;\mycalm ) \cap \partial\mycalu$ and its image under
  $\psi$.}
 \label{condHE.fig}
\end{figure}
Under those conditions the area theorem does again hold.

Several other proposals how to modify the WASE conditions of
\cite{HE} to obtain sufficient control of the space--times at hand
have been proposed in
\cite{Newman:coscen,Krolak:erpfsps,ClarkedeFelice2}.

A completely different, and considerably simpler, treatment of the
question of regularity of Scri needed for the area theorem, has
been proposed in~\cite{ChDGH}: A point $q$ in a set $A\subset B$
is said to be a {\em past} point of $A$ with respect to $B$ if
${J^-(q;B)}\cap {A}=\{q\}$. One then says that $\scrip$ is
$\cH$--\emph{regular} if there exists a neighborhood $\cal O$ of
$\cH$ such that for every compact set $C\subset\cal O$ satisfying
$I^+(C;\bar M) \cap \Scri^+ \ne \emptyset$ there exists a past
point \emph{with respect to $\Scri^+$} in $\partial I^+(C; \bar
M)\cap \Scri^+$.  The condition of $\cH$--{regularity} of $\scrip$
is {precisely} what is needed for the arguments of the area
theorem to go through.  This condition is somewhat related to the
\emph{$i^0$-avoidance condition} used by Galloway and
Woolgar~\cite{galloway:woolgar}.

Yet another approach has been advocated by Wald~\cite{Wald:book}:
there one considers globally hyperbolic completions in a set-up
that {includes $i^0$}. While the conditions of~\cite{Wald:book} do
lead to a coherent set-up for the validity of the area theorem,
they introduce several new difficulties related to the low
differentiability of the conformally rescaled metric at $i^0$, not
to mention the question of mere existence of such completions.
This can actually be relaxed to the related weaker requirement of
global hyperbolicity of $\mcM \cup \scrip$ (in the sense of a
manifold with boundary) which enforces $\cH$--{regularity} of
$\scrip$~\cite{ChDGH}, so that the area theorem again holds.

\chindex{Scri!regularity conditions|)} \bhindex{gravitational!Scri
  based|)}
\section{Horizons}
\chindex{horizons!differentiability of|(}

A key notion related to the concept of a black hole is that of an
\emph{event horizon}, \bel{eh1} \mcE:= \partial \mcB\;.\ee
(Actually, when $\mcE $ is not connected some care is required for
a useful definition, but this does not affect the discussion that
follows.) Event horizons are a special case of a family of objects
called \emph{future horizons}: by definition, these are closed
topological hypersurfaces $\mcH$ threaded by null geodesics,
called {\em generators}, with no \emph{future end} points, and
possibly with \emph{past end} points. At the latter,
differentiability of $\mcH$ breaks down in general; a necessary
and sufficient condition for this breakdown has been given
in~\cite{BK2}.  It seems that most authors have been taking for
granted that horizons are nice, piecewise smooth hypersurfaces.
This is, however, not the case, and examples of {\em nowhere}
$C^1$ horizons have been constructed in~\cite{ChGalloway}.
Further, {nowhere} $C^1$ horizons are {generic} in the class of
convex horizons in Minkowski space--time~\cite{BKK}. This leads to
various difficulties when trying to study their structure, e.g.,
attempting to prove results such as the area theorem discussed in
Sect.~\ref{Scrc}, compare Theorem~\ref{Tqlbharea} below. Horizons
are always {\em locally Lipschitz-continuous} topological
hypersurfaces~\cite{PenroseDiffTopo} (compare Corollary~\ref{C1}
below), i.e., they are locally graphs of functions satisfying
$$|\phi(x)-\phi(y)|\le C |x-y|\;.$$
In spite of the potential low differentiability, the usual optical
scalars can be defined {almost
  everywhere} in an \emph{Alexandrov sense}, as follows: A function
  is said to be \emph{semi--convex} if it is the sum of a $C^2$
  function and of a convex function. It has been shown
  in~\cite{ChDGH}\footnote{Some further results concerning
    differentiability properties of horizons can be found
    in~\cite{BK2,Chendpoints,ChFGH}.}  that future horizons are
  semi-convex.  We give an alternative proof of this result in an
  appendix: the new argument has some interest on its own, as it
  proceeds via a variational principle for horizons, which could be
  useful for numerical applications.  The interest of the
  semi-convexity property relies on the following theorem of
  Alexandrov:
\begin{ptctheorem}\emph{\bf
    (Alexandrov~\cite[Appendix~E]{FlemingSoner})}\chindex{Alexandrov's
    theorem}
  \label{Tdiff} Semi--convex functions $f:B\to\R$ are ``twice--differentiable''
  \emph{almost everywhere} in the following sense: there exists a set
  $\BA$ with full measure in $B$ such that
  \beaa&\forall\ x\in \BA\ \exists 
  \ Q\in (\R^p)^*\otimes (\R^p)^*\ \mbox{\rm such that}\ \forall \
  y\in B \phantom{df(x)(y-x)=}&\nonumber\\ &\quad f(y)-f(x) -
  df(x)(y-x)=Q(x-y,x-y)+ r_2(x,y)\ ,& \label{alex2}\eeaa with
  $r_2(x,y)=o(|x-y|^2)$. The symmetric quadratic form $Q$ above will
  be denoted by $\frac{1}{2}D^2f(x)$, and will be called the {\em
    second Alexandrov derivative} of $f$ at $x$.
\end{ptctheorem}

Points in $\BA$ will be referred to as \emph{Alexandrov points} of
$B$.

One can now use Theorem~\ref{Tdiff} to define the equivalent to
the usual divergence $\theta$ of the horizon, by writing $\cH$
locally as the graph of a function $f$, and using the second
Alexandrov derivatives of $f$ to define (almost everywhere on
$\cH$) the divergence $\tA$ of the horizon.  More precisely, let
$p=(t=f(q),q)$ be an Alexandrov point of $\cH$ and let $e_i$,
$i=1,2,3$ a basis of $T_p\cH$ such that
\begin{eqnarray*}
  & e_3=K_\mu(p)dx^\mu= -dt+df(q)\;, \ g(e_1,e_1)=(e_{2},e_{2})=1\; ,
  \ g(e_1,e_2)=0
  \;. &\end{eqnarray*} One then sets \begin{eqnarray*}& \nabla_i K_j =
  D^2_{ij}f - \Gamma^{\mu}_{ij}
K_\mu \ , 
\qquad \tA  =(e_1^i e_1^j +e_{2}^i e_{2}^j) \nabla_i K_j \ .
&\label{theta}
\end{eqnarray*}
{where the Hessian matrix $D^2_{ij}f$ above is that of second
  Alexandrov derivatives. This reduces to the standard definition of
  $\theta$ when $f\in C^2$.}

Several results of causality theory go through with $\theta$
replaced by $\tA$, though the standard arguments sometimes have to
be replaced by completely different ones, the reader is referred
to~\cite{ChDGH} for details. In particular the Raychaudhuri
equation, as well as the remaining optical equations, hold {on
almost all} generators of $\mcH$.

Somewhat surprisingly, event horizons in {smooth, stationary,
globally
  hyperbolic, asymptotically flat} space-times satisfying the null
energy condition are always {smooth} null hypersurfaces, analytic
if the metric is analytic; this is a corollary of the area theorem
of~\cite{ChDGH}.

\chindex{horizons!differentiability of|)}
\section{{Apparent horizons}}\label{Sah}

\chindex{horizons!apparent} In spite of their name, apparent
horizons are \emph{not} horizons. They are usually~\cite{HE}
defined on spacelike hypersurfaces $\hyp\subset\cM$, as follows:
let {$\Omega\subset \cM$} be the {set covered by {\em (future)
trapped
    surfaces}:\chindex{trapped surface} by definition, those are
  compact, boundaryless, smooth surfaces $S\subset \hyp$ with the
  property that } \bel{trsur} \left.\theta(S):= \lambda- (g^{ij}-
  n^in^j) K_{ij} < 0\right.\;,\ee {where $\lambda$ is the (outwards)
  mean extrinsic curvature of $S$ in $\hyp$, $n^i$ is the outwards pointing
  unit normal to $S$ in $\hyp$, and $K_{ij}$ is the extrinsic
  curvature of $\hyp$ in $\cM$.} There is no reason for $\Omega$ to be
nonempty in general; on the other hand, in appropriately censored
space-times, a non-empty $\Omega$ implies the existence of a black
hole region. Hawking and Ellis define the apparent horizon $\cA$
as \be\label{HE1}\cA:=\partial\Omega\;,\ee and argue that \bel{tv}
\theta(\cA)=0\;.\ee Their argument is correct if one {assumes}
that $\cA$ is $C^2$. However, $\cA$ could be a priori a very rough
set, with $\theta$ {not even defined}, in which case the arguments
of~\cite{HE} do not apply. Probably the simplest way out is to use
\eq{tv} as the definition of apparent horizon, and forget about
$\Omega$. The existence of a $C^2$ compact, boundaryless surface
satisfying \eq{tv} does again imply the existence of a black hole
in a Scri framework, assuming appropriate causal properties of
$\scrip$.

We note that if $K_{ij}=0$, then (\ref{HE1}) is the equation for a
minimal surface. Now, in the course of their proof of the Penrose
inequality,
G.~Huisken and T.~Ilmanen~\cite{HI2} prove that 
{the outermost minimal (in the sense of calculus of variations)
  surface is always smooth}, which supports the validity of \eq{tv} at
least for the outermost component of $\cA$. {However, they provide
  examples which show that this outermost minimal surface might not be
  the boundary of $\Omega$ in general.} This illustrates another
deficiency of the definition \eq{HE1}. Some partial results
concerning the differentiability properties of $\cA$ have been
obtained by M.~Kriele and S.~Hayward in~\cite{KrieleHayward}.
R.~Howard and J.~Fu have recently shown~\cite{HowardFu} that
$\partial \Omega$ satisfies (\ref{HE1}) in a {\em viscosity sense}
-- this is defined as follows: Let $D$ be an open set, and let
$p\in \partial D$. Then $U$ is an {inner} ({outer}) support domain
if $U$ is an open {subset} {of $D$} ({of $\complement D$}),
$\partial U$ is a $C^2$ hypersurface and $p\in
\partial U$. One says that {$\theta(\partial D) \le 0$} (respectively
{$\theta(\partial D) \ge 0$}) in the viscosity sense if for any
point $p\in\partial D$ and for any {inner} support ({outer
support}) domain we have
$$
{\theta(\partial U)(p)\le 0} \qquad (\textrm{respectively }\
{\theta(\partial U)(p)\ge 0} ) \;. $$ {This coincides with the
usual
  inequality {$ {\theta\le 0}$} ({$ {\theta\ge 0}$) on any subset of
    $\partial D$ which is $C^2$.}} Finally $\theta(\partial D)=0$ in
the viscosity sense if for any point $p\in\partial D$ we have both
{$\theta(\partial D) \le 0$} in the viscosity sense, and
{$\theta(\partial D) \ge 0$} in the viscosity sense.

It is tempting to conjecture the following: if ${ {\widehat
\Omega}}$ is the set covered by smooth compact $S$'s satisfying
$$\left.\theta(S):= \lambda+ (g^{ij}- n^in^j) K_{ij}\; {\le}\;
  0\right.\;,$$
then any outermost connected component of $\partial\widehat\Omega$
is smooth, separating, and satisfies \eq{HE1}.  When $K_{ij}=0$
this statement is known to be true by the already mentioned result
of G.~Huisken and T.~Ilmanen.

We refer the reader to~\cite{Alcubierre:1998rq,Gundlach:1998us}
and references therein for a numerical treatment of apparent
horizons.

\section{Classification of stationary solutions (``No hair
  theorems")} \chindex{no hair
  theorems|(}\bhindex{gravitational!uniqueness theorems|(}

Uniqueness theorems for stationary solutions have been one of the
central fields of research in the theory of black holes, the
reader is referred to~\cite{ChAscona,Chnohair} and references
therein for a review, as well as lists of open problems; cf.\
also~\cite{Heusler:book,Heusler:living}. Actually the perspective
here is somewhat larger, as one is interested in classifying all
solutions, not only those that contain black holes. Since the time
of writing of \cite{ChAscona,Chnohair} some progress has been
made, both for zero and non-zero cosmological constant; it is
convenient to discuss various cases separately. Throughout this
section, \emph{stationary} means that there exists an action of
$\R$ on $\mcM$ by isometries, with the orbits of the group being
timelike sufficiently far away in an asymptotic region;
\emph{strictly stationary} means that the orbits are timelike
everywhere. Stationary black hole solutions are never strictly
stationary.

\subsubsection{$\Lambda=0$:}
The following fundamental theorem has been recently proved by
M.~Anderson: \chindex{Anderson's rigidity theorem}
\begin{ptctheorem}[Anderson~\cite{manderson:static}] The only geodesically complete strictly stationary
  solutions of the vacuum Einstein equations which do not contain
  closed timelike curves are the Minkowski space-time and quotients
  thereof.
\end{ptctheorem}

This is a milestone result in general relativity, involving only
the natural geometric conditions of completeness together with
absence of closed timelike curves. This should be contrasted with
the version of this result originating in Lichnerowicz's work, the
state-of-the art form of which, as based on previous techniques,
assumes {\em in
  addition} global hyperbolicity, {\em and} asymptotic flatness, {\em
  and} existence of a compact interior. Last but not least, the paper
introduces techniques which have not been used in mathematical
general relativity so far, and which are likely to be useful tools
in future work.

In the black hole case, the only significant progress made since
the review paper~\cite{ChAscona} is for {\em static} black holes:
In~\cite{Chstatic} the classification of static vacuum black holes
which contain an asymptotically flat spacelike hypersurface with
compact interior has been, essentially,\footnote{Some comments
about
  the qualification ``essentially" are in order: uniqueness theorems
  usually proceed in two steps, the first being the reduction of the
  problem to a Riemannian one, the second consisting of a uniqueness
  theorem for the Riemannian problem. The latter part of the proof
  in~\cite{Chstatic} seems to be optimal; one could consider improving
  the former, see~\cite{Chstatic} for a detailed discussion.}
finished. This problem has a long history, starting with the
pioneering work of Israel~\cite{Israel:vacuum}.  The most complete
result existing in the literature prior to~\cite{Chstatic} was
that of Bunting and Masood-ul-Alam~\cite{bunting:masood} who
showed, roughly speaking, that all appropriately regular such
black holes which {\em
  do not contain degenerate horizons} belong to the Schwarzschild
family. In~\cite{Chstatic} the condition of non-degeneracy of the
event horizon is removed, showing that the Schwarzschild black
holes exhaust the family of all appropriately regular black hole
space-times.  The proof requires an extension of the positive mass
theorem which applies to asymptotically flat {\em complete}
Riemannian manifolds, proved in~\cite{BartnikChrusciel}.  The
paper \cite{Chstaticelvac} contains an improvement of similar
previous results concerning the electro-vacuum black holes, under
the restrictive condition that all degenerate components of the
black hole carry charges of the same sign. This seems a reasonable
condition from a physical point of view: opposite charges would
attract, which added to the gravitational attraction should
prevent the configuration from being static. However, a proper
mathematical treatment which would transform this argument into a
proof is still missing.

\subsubsection{$\Lambda > 0$:} The natural boundary conditions
for stationary solutions with $\Lambda > 0$ are the ``no boundary
ones": one considers globally hyperbolic space-times containing
compact, boundaryless spatial surfaces. In this case there is no
asymptotic region, and {\em stationarity} is defined by the
requirement that the set of points at which the Killing vector
field is timelike is non-empty. In~\cite{LafontaineRozoy}
J.~Lafontaine and L.~Rozoy have announced a complete
classification of static space-times with a compact (boundaryless)
totally geodesic spacelike hypersurface $\hyp$, under the
assumption that the Lorentzian norm $\sqrt{-g_{\mu\nu} X^\mu
X^\nu}$ of the Killing vector field is a Morse-Bott function, and
that of analyticity\footnote{This is a
  condition on the analyticity of the metric at the set of points at
  which the Killing vector vanishes or becomes null, since the metric
  is necessarily analytic elsewhere.} of the metric on $\hyp$. The
hypothesis that $\hyp$ is totally geodesic is the usual hypothesis
of staticity. This is the first result of such generality in this
context. The proof is an extension of an argument of R.~Beig and
W.~Simon~\cite{BeigSimon3} done in a related context, and proceeds
by proving that the metric on $\hyp$ must be conformally flat; one
can then conclude using~\cite{Lafontaine}.

Uniqueness results under the completely different hypothesis of
existence of a conformal completion have been previously
established by H.~Friedrich~\cite{FriedrichdS} and by
Boucher~\cite{Boucher}.

\subsubsection{$\Lambda<0$:}

Our understanding of the class of stationary space-times with a
negative cosmological constant is much poorer than that of the
$\Lambda=0$ ones. It seems that the only results available so far
concern static space-times. To start with, the question of
boundary conditions which should be imposed in the asymptotic
region are far from being understood, a beginning of a systematic
analysis of this question can be found in~\cite{ChruscielSimon}. A
uniqueness argument for the de Sitter metric in the strictly
stationary case has been proposed in~\cite{BGH}. It is only
recently that complete proofs of the required positive energy
theorem have been given~\cite{Wang,Zhang:hpet,ChHerzlich}; the
results in~\cite{ChHerzlich} do actually lead to a uniqueness
result under much weaker asymptotic conditions than originally
suggested~\cite{BGH}. It should be pointed out that the nature of
the corresponding Riemannian problem is completely different from
that of the asymptotically flat case: here one needs to analyze
the set of solutions of the equations for a Riemannian metric $g$
and a function $V$ on a three dimensional manifold $\hyp$,
\begin{eqnarray}
\label{f1bc}& \Delta V  =  - \Lambda V\; ,& \\& \label{f2bc}
R_{ij}  =  V^{-1} D_{i} D_{j} V + \Lambda g_{ij}\;, &
\end {eqnarray}
and it is customary to assume that $(\hyp,g)$ can be conformally
compactified. Now, the uniqueness theorems mentioned require that
$V^{-2}g$ extends by continuity to a smooth metric on the
conformal boundary, with constant scalar curvature there. No
reasons are known why this should be the case, and it is
conceivable that more general static solutions would exist which
do not satisfy this condition. It would be of interest to
construct such solutions, or to prove their non-existence.

The topology of the boundary of the conformally compactifiable
solutions of \eqs{f1bc}{f2bc} does not have to be spherical, as is
the case for the anti-de Sitter metric, while the uniqueness
result mentioned above assumes that the boundary at infinity
$\pihyp$ is a sphere. When $\pihyp$ is a torus,  a strictly static
solution has been found by Horowitz and Myers
\cite{HorowitzMyers}. The metric in $n+1\geq 4$ spacetime
dimensions is
\begin{equation}
ds^2 = -r^2dt^2 + \frac{1}{V(r)}dr^2 + V(r)d\phi^2 +r^2
\sum\limits_{i=1}^{n-2}(dy^i)^2 \;, \label{Intro1}
\end{equation}
where $V(r)= \frac{r^2}{\ell^2} \left ( 1-\frac{\ro^{n}}{r^{n}}
\right )$, $\ell^2=-\frac{n(n-1)}{2\Lambda}$, and $\ro$ is a
constant. Regularity demands that $\phi$ be identified with period
$\beta_0=\frac{4\pi\ell^2}{n\ro}$. The periods of the $y^i$'s are
arbitrary. The space-time metric is ``asymptotically locally
anti-de Sitter'' with boundary at conformal infinity  foliated by
spacelike $(n-1)$-tori. The time slices of space-time itself, when
conformally completed, have topology $D^2 \times T^{n-2}$, which
in dimension 3+1 is a solid torus. Under a convexity condition on
the conformal boundary, G.~Galloway, S.~Surya and E.~Woolgar have
proved a uniqueness theorem for the corresponding
metrics~\cite{Galloway:2001uv}.

The above mentioned results concern the strictly static case, and
the question of classification of  black holes with a Killing
vector which is not timelike everywhere is essentially open.
In~\cite{ChruscielSimon} an argument has been presented which
shifts the problem of proving uniqueness of a class of such black
holes to that of proving a Penrose-type inequality\footnote{A
version of such an inequality has been originally proposed by
Gibbons~\cite{GibbonsGPI}; the inequality proposed there is not
quite correct, see~\cite{ChruscielSimon}.} for conformally
compactifiable Riemannian manifolds with minimal boundary, and
with scalar curvature bounded from below by a negative constant.
While there is some hope that some such results could be proved by
extending the arguments of~\cite{HI2}, this remains to be seen.

\chindex{no hair theorems|)}\bhindex{gravitational!uniqueness
theorems|)}

\section{Black holes without Scri}

There has been considerable progress in the numerical analysis of
black hole solutions of Einstein's equations; here one of the
objectives is to write a stable code which would solve the full
four dimensional Einstein equations, with initial data containing
a non-connected black hole region that eventually merges into a
connected one. One wishes to be able to consider initial data
which do not possess any symmetries, and which have various
parameters -- such as the masses of the individual black holes,
their angular momenta, as well as  distances between them -- which
can be varied in significant ranges. Finally one wishes the code
to run to a stage where the solution settles to a state close to
equilibrium. The challenge then is to calculate the gravitational
wave forms for each set of parameters, which could then be used in
the gravitational wave observatories to determine the parameters
of the collapsing black holes out of the observations made. This
program has been being undertaken for years by several groups of
researchers, with steady progress being
made~\cite{Bishop:1997ik,Bartnik:NQSURL,Gomez:1998uj,%
Alcubierre:2001vm,Alcubierre:2000ke,Kidder:2001tz,Grandclement:2001ed,%
Brandt:2000yp,Lehner:2001wq}.\footnote{Some spectacular
visualizations of the calculations performed can be found at the
URL \url{http://jean-luc.aei.mpg.de/NCSA1999/GrazingBlackHoles}}

There is a fundamental difficulty above, of deciding whether or
not one is dealing indeed with the desired black hole initial
data: the definition \eq{sbh1} of a black hole requires a
conformal boundary $\scri$ satisfying some -- if not all --
properties discussed in \Sect{Scrc}. Clearly there is no way of
ensuring those requirements in a calculation performed on a finite
space-time grid.\footnote{The conformal approach developed by
 Friedrich (cf., e.g., \cite{Friedrich:Pune} and references therein, as well
 as S.~Husa's and J.~Frauendiener's contributions to this volume)
 provides an ideal numerical framework
 for studying gravitational radiation in situations where the extended space-time
 is smoothly conformally compactifiable across $i^+$, since then
 one can hope that the code will be able to ``calculate Scri"
 globally to the future of the initial hyperboloidal hypersurface.
 It is not clear whether  a conformal approach could provide more
 information than the non-conformal ones
 when $i^+$ is itself a singularity of
 the conformally rescaled equations, as is the case for black
 holes.}

In practice what one does is to set up initial data on a finite
grid so that the region near the boundary is close to flat (in the
conformal approach the whole asymptotically flat region is covered
by the numerical grid, and does not need to be near the boundary
of the numerical grid; this distinction does not affect the
discussion here). Then one evolves the initial data as long as the
code allows. The gravitational waves emitted by the system are
then extracted out of the metric near the boundary of the grid.
Now, our understanding of energy emitted by gravitational
radiation is essentially based on an analysis of the metric in an
asymptotic region where $g$ is nearly flat.  In order to recover
useful information out of the numerical data it is thus necessary
for the metric near the boundary of the grid to remain close to a
flat one. If we want to be sure that the information extracted
contains all the essential dynamical information about the system,
the metric near the boundary of the grid should quiet down to an
almost stationary state as time evolves. Now, it is
straightforward to set-up a mathematical framework to describe
such situations without having to invoke conformal completions,
this is done in the next section.

\subsection{Naive black holes}\bhindex{gravitational!naive}
\label{Snbh}
 Consider a {\em globally hyperbolic} space-time $\cM$ which contains a region  covered by
  coordinates  $(t,x^i)$ with ranges
   \bel{nbh1} r:=\sqrt{\sum_i (x^i)^2} \ge R_0\;,\qquad  T_0-R_0 +r\le t
<\infty \;,\ee such that the metric $g$ satisfies there \bel{nbh2}
|g_{\mu\nu}-\eta_{\mu\nu}|\le C_1r^{-\alpha}\le C_2\;, \quad
\alpha >0\;,\ee for some positive constants $C_1,C_2,\alpha$;
clearly $C_2$ can be chosen to be less than or equal to $
C_1R_0^{-\alpha}$. Making $R_0$ larger one can thus make $C_2$ as
small as desired, e.g. \bel{nbh3}C_2= 10^{-2}\;,\ee which is a
convenient number in dimension $3+1$ to guarantee that objects
algebraically constructed out of $g$ (such as $g^{\mu\nu},
\sqrt{\det g}$) are well controlled; \eq{nbh3} is certainly not
optimal, and any other number suitable for the purposes at hand
would do. To be able to prove theorems about such space-times one
would need to impose some further, perhaps not necessarily
uniform, decay conditions on a finite number of derivatives of
$g$; there are various possibilities here, but we shall ignore
this for the moment. Then one can define the {\em exterior region}
$\cMext$, the
 {\em black hole region} $\mycal B$ and the
 {\em future event horizon} $\event$ as
\bel{nbh4.0}\cMext := \cup_{\tau\ge T_0} J^-({\cS_{\tau,R_0}})=
J^-(\cup_{\tau\ge T_0}{\cS_{\tau,R_0}}) \;,\ee \bel{nbh4}\quad \cB
:= \cM \setminus \cMext\;,\quad \event :=\partial \cB\;,\ee where
\bel{nbh5} \cS_{\tau,R_0}:= \{t=\tau, r=R_0\}\;.\ee We will refer
to the definition \eqs{nbh1}{nbh5} as that of a \emph{naive} black
hole.

 In the setup of
\Eqs{nbh1}{nbh5} an arbitrarily chosen $R_0$ has been used; for
this definition to make sense $\cB$ so defined should \emph{not}
depend upon this choice. This is indeed the case, as can be seen
as follows:
\begin{ptcproposition}\label{Pnbh1}
  Let $\mcO_a\subset\R^3\setminus B(0,R_0)$, $a=1,2$, and let
  $\mcU_a\subset\mcM$ be of the form $\{(t\ge T_0-R_0 +r(\vec x),\vec
  x)\;, \vec x \in \mcO_a\}$ in the coordinate system of \eq{nbh2}.
  Then
  $$I^-(\mcU_1)=J^-(\mcU_1) =I^-(\mcU_2)= J^-(\mcU_2)\;.$$
\end{ptcproposition}
\begin{proof}
  If $\Gamma$ is a future directed causal path from $p\in \mcM$ to
  $q=(t,\vec x) \in \mcU_1$, then the path obtained by concatenating
  $\Gamma$ with the path $[0,1]\ni s\to (t(s):=t+s,\vec x(s):=\vec x)$
  is a causal path which is not a null geodesic, hence can be deformed
  to a timelike path from $p$ to $(t+1,\vec x)\in \mcU_1$. It follows
  that $I^-(\mcU_1)=J^-(\mcU_1)$; clearly the same holds for $\mcU_2$.
  Next, let $\vec x_a \in \mcO_a$, and let
  $\gamma:[0,1]\to\R^3\setminus B(0,R_0)$ be any differentiable path
  such that $\gamma(0)=\vec x_1$ and $\gamma(1)=\vec x_2$. Then for
  any $t_0\ge T_0-R_0 +r(\vec x_1)$ the causal curve $[0,1]\ni
  s\to\Gamma(s)=\left( t:=Cs+t_0,\vec x(s):=\gamma(s)\right)$ will be
  causal for the metric $g$ by \eq{nbh2} if the constant $C$ is chosen
  large enough, with a similar result holding when $\vec x_1$ is
  interchanged with $\vec x_2$. The equality $I^-(\mcU_1)
  =I^-(\mcU_2)$ easily follows from this observation. \qed
\end{proof}
Summarizing, Prop.~\ref{Pnbh1} shows that there are many possible
{equivalent definitions} of $\mcM_{\ext}$: in \eq{nbh4.0} one can
replace $ J^-({\cS_{\tau,R_0}})$ by $ J^-({\cS_{\tau,R_1}})$ for
any $R_1\ge R_0$, but also simply by $J^-((t+\tau,q))$, for any
$p=(t,q)\in \mcM$ which belongs to the region covered by the
coordinate system $(t,x^i)$.

The following remarks concerning the definition \Eqs{nbh4.0}{nbh4}
are in order:
\begin{itemize}
\item For vacuum, stationary, asymptotically flat space-times the
  definition is {equivalent} to the usual one with $\scri$
  \cite[Footnote~7, p.~572]{ChWald}; here the results of
  \cite{Dain:2001kn,Damour:schmidt} are used. However, one does not
  expect the existence of a smooth $\scrip$ to follow from
  \eq{nbh1}-\eq{nbh2} in general.
\item Suppose that $\mcM$ admits a conformal completion in the sense
  defined at the beginning of \Sect{Ssbh}, and that $\Scri$ is {\em
    semi-complete to the future}, in the sense that the
  Geroch-Horowitz condition of the beginning of \Sect{Scrc} holds with
  $\scri\approx \R\times S^2$ there replaced by $\scri\supset
  \R^+\times S^2$. Then for any finite interval $[T_0,T_1]$ there
  exists $R_0(T_0,T_1)$ and a coordinate system satisfying \eq{nbh2}
  and covering a set $r\ge R_0(T_0,T_1)$, $T_0-R_0 \le t \le T_1-R_0$.
  This follows from the Tamburino-Winicour construction of Bondi
  coordinates $(u,r,\theta,\varphi)$ near
  $\scrip$~\cite{TamburinoWinicour}, followed by the introduction of
  the usual Minkowskian coordinates $t=u+r$,
  $x=r\sin\theta\cos\varphi$, {\em etc.} The problem is that
  $R(T_1,T_2)$ could shrink to zero as $T_2$ goes to infinity. Thus,
  when $\scrip$ exists, conditions \Eqs{nbh1}{nbh2} are {\em
    uniformity} conditions on $\scrip$ to the future -- the metric
  remains uniformly controlled on a uniform neighbourhood of $\scrip$
  as the retarded time goes to infinity.
\item It should not be too difficult to check whether or not the
  future geodesically complete space-times of
  Friedrich~\cite{friedrich:cauchy,FriedrichSchmidt} and of
  Christodoulou and Klainerman~\cite{Ch-Kl}, or the Robinson-Trautman
  black-holes~\cite{ChRT2}, admit coordinate systems satisfying
  \eqs{nbh1}{nbh2}.
\end{itemize}

It is not clear if asymptotically flat space-times in which no
such control is available do exist at all; in fact, it is tempting
to formulate the following version of the {\em (weak) cosmic
censorship} conjecture:
 \begin{quote}The maximal globally
   hyperbolic development of generic\footnote{The examples constructed
     by Christodoulou~\cite{demetrios:scalar10} with spherically
     symmetric gravitating scalar fields suggest that the genericity
     condition is unavoidable, though no corresponding vacuum examples
     are known.}, asymptotically flat, vacuum initial data contains a
   region with coordinates satisfying \eq{nbh1}-\eq{nbh2}.
\end{quote} Whatever the status of this conjecture, one can hardly
envisage numerical simulations leading to the calculation of an
essential fraction of the total energy radiated away in
space-times in which some uniformity conditions do not hold.

\subsection{Quasi-local black holes}\label{Sqlbh}

\bhindex{gravitational!quasi-local} As already argued, the naive
approach of the previous section should be more convenient for
numerical simulations of black hole space-times, as compared to
the usual one based on Scri. It appears to be even more convenient
to have a framework in which all the issues are localized in
space; we wish to suggest such a framework here. When numerically
modeling an asymptotically flat space-time, whether in a conformal
or a direct approach, a typical numerical grid will contain large
spheres S(R) on which the metric is nearly flat, so that an
inequality such as \eq{nbh2}-\eq{nbh3} will hold in a
neighbourhood of S(R). On slices $t=\const$ the condition
\eq{nbh2} is usually complemented with a fall-off condition on the
derivatives of the metric \bel{qlbh1} |\partial_\sigma
g_{\mu\nu}|\le C r^{-\alpha-1}\;, \ee however condition \eq{qlbh1}
is inadequate in the radiation regime, where retarded time
derivatives of the metric are not expected to fall-off faster than
$r^{-1}$. It turns out that there is a condition on derivatives of
the metric in null directions which is fulfilled at large distance
both in spacelike and in null directions: let $K_a$, $a=1,2$, be
null future pointing vector fields on $S(R)$ orthogonal to $S(R)$,
with $K_1$ inwards pointing and $K_2$ -- outwards pointing; these
vector fields are unique up to scaling. Let $\theta_a$ denote the
associated null second fundamental forms defined as
\bel{qlbh1.1}\forall \ X, Y\in TS(R)\quad \chi_a(X,Y):=
g(\nabla_XK_a,Y)\;.\ee It can be checked, e.g. using the
asymptotic expansions for the connection coefficients near
$\scrip$ from~\cite[Appendix~C]{CJK}, that $\chi_1$ is negative
definite and $\chi_2$ is positive definite for Bondi spheres
$S(R)$ sufficiently close to $\scrip$; similarly for $\scri^-$.
This property is not affected by the rescaling freedom at hand.
Following G.~Galloway~\cite{Galloway:fitopology}, a
two-dimensional spacelike submanifold of a four-dimensional
space-time will be called \emph{\wnc} if $\chi_1$ is semi-positive
definite, with the trace of $\chi_2$ negative.\footnote{Galloway
defines \emph{null convexity} through the requirement of positive
definiteness of $\chi_1$ and negative definiteness of $\chi_2$.
However, he points out himself~\cite[p.~1472]{Galloway:fitopology}
that the weak null convexity as defined above suffices for his
arguments to go through.} The null convexity condition is easily
verified for sufficiently large spheres in a region asymptotically
flat in the sense of \eq{qlbh1}. It does also hold for large
spheres in a large class of space-times with negative cosmological
constant. The null convexity condition is then the condition which
we propose as a starting point to defining ``quasi-local" black
holes and horizons. The point is that several of the usual
properties of black holes carry over to the \wnc\ setting. In
retrospect, the situation can be summarized as follows: the usual
theory of Scri based black holes exploits the existence of
conjugate points on appropriate null geodesics whenever those are
complete to the future; this completeness is guaranteed by the
fact that the conformal factor goes to zero at the conformal
boundary at an appropriate rate. Galloway's discovery
in~\cite{Galloway:fitopology} is that weak null convexity of large
spheres near Scri provides a second, in principle completely
independent, mechanism to produce the needed focusing behaviour.

Throughout this section we will consider a globally hyperbolic
space-time $(\mcM,g)$ with time function $t$. Let
$\mcT\subset\mcM$ be a finite union of connected timelike
hypersurfaces $\mcT_\alpha$ in $\mcM$, we set \bel{qlbh2}
\hyp_\tau:= \{t=\tau\}\;,\quad\mcT(\tau):=
\mcT\cap\hyp_\tau\;,\quad\mcT_\alpha(\tau):=
\mcT_\alpha\cap\hyp_\tau\;.\ee For further purposes anything that
happens on the exterior side of $\mcT$ is completely irrelevant,
so it is convenient to think of $\mcT$ as a boundary of $\mcM$;
global hyperbolicity should then be understood in the sense that
$(\bmcM:= \mcM\cup\mcT,g)$ is strongly causal, and that
$J^+(p;\bmcM)\cap J^-(q;\bmcM)$ is compact in $\bmcM$ for all
$p,q\in \bmcM$. Recall that the null convergence condition is the
requirement that \bel{ncc} \Ric(X,X)\ge 0 \quad \textrm{for all} \
X\in TM\;.\ee We have the following \emph{topological censorship}
theorem for \wnc\ timelike boundaries:

\begin{ptctheorem}[Galloway~\cite{Galloway:fitopology}]\label{Tqlbh1}
  Suppose that a globally hyperbolic space-time $(\bmcM,g)$ satisfying
  the null convergence condition \eq{ncc} has a timelike boundary
  $\mcT=\cup_{\alpha=1}^I \mcT_\alpha$ and a time function $t$ such
  that the level sets of $t$ are Cauchy surfaces, with each section
  $\mcT(\tau)$ of $\mcT$ being null convex. Then distinct
  $\mcT_\alpha$'s cannot communicate with each other:
  $$
  \alpha\ne\beta \qquad J^+(\mcT_\alpha)\cap
  J^-(\mcT_\beta)=\emptyset\;.$$
\end{ptctheorem}

 As is well known,
topological censorship implies constraints on the topology:

\begin{ptctheorem}[Galloway~\cite{Galloway:fitopology}]
\label{Tqlbh2} Under the hypotheses of Theorem~\ref{Tqlbh1}
suppose further that the cross-sections $\mcT_\alpha(\tau)$ of
$\mcT_\alpha$ have spherical topology.\footnote{The reader is
referred
  to~\cite{Galloway:1999br} and references therein for results without
  the hypothesis of spherical topology. The results there, presented
  in a Scri setting, generalize immediately to the \wnc\ one.} Then
the \emph{$\alpha$-domain of outer communication} \bel{ddoc}
\langle\langle\mcT_\alpha\rangle\rangle:= J^+(\mcT_\alpha)\cap
J^-(\mcT_\alpha)\ee is simply connected.
\end{ptctheorem}

It follows in particular from Theorem~\ref{Tqlbh2} that $\bcM$ can
be replaced by a subset thereof such that $\mcT$ is connected in
the new space-time, with all essential properties relevant for the
discussion in the remainder of this section being unaffected by
that replacement. We shall not do that, to avoid a lengthy
discussion of which properties are relevant and which are not, but
the reader should keep in mind that the hypothesis of
connectedness of $\mcT$ can indeed be done without any loss of
generality for most purposes.

 We define the \emph{quasi-local black hole region}
$\mcB_{\mcT_\alpha}$ and the \emph{quasi-local event
horizon}\chindex{horizons!quasi-local} $\mcE_{\mcT_\alpha}$
associated with the hypersurface ${\mcT_\alpha}$ by \bel{qlbh3}
\mcB_{\mcT_\alpha}:= \mcM\setminus J^-(\mcT_\alpha)\;, \qquad
\mcE_{\mcT_\alpha}:=
\partial \mcB_{\mcT_\alpha} \;.\ee
 \emph{If} $\mcT$ is the hypersurface $\cup_{\tau\ge
T_0}{\cS_{\tau,R_0}}$ of Section~\ref{Snbh} then the resulting
black hole region coincides with that defined in \eq{nbh4}, hence
does not depend upon the choice of $R_0$ by
Proposition~\ref{Pnbh1}; however, $\mcB_{\mcT_\alpha}$ might
depend upon the chosen family of observers ${\mcT_\alpha}$ in
general. It is certainly necessary to impose some further
conditions on $\mcT$ to reduce this dependence. A possible
condition, suggested by the geometry of the large coordinate
spheres considered in the previous section, could be that the
light-cones of the induced metric on $\mcT$ are uniformly
controlled both from outside and inside by those of  two static,
future complete reference metrics on $\mcT$. However, neither the
results above, nor the results that follow, do require that
condition.

The Scri-equivalents of
Theorem~\ref{Tqlbh2}~\cite{GallowayBrowdy,Jacobson:venkatarami,%
ChWald,galloway-topology,galloway:woolgar,Galloway:1999br} allow
one to control the topology of ``good" sections of the horizon,
and for the standard stationary black-holes this does lead to the
usual $S^2\times \R$ topology of the horizon~\cite{HE,ChWald}. In
particular, in stationary, asymptotically flat, appropriately
regular space-times the intersection of a partial Cauchy
hypersurface with an event horizon will necessary be a finite
union of spheres. In general space-times such intersections do not
even need to be manifolds: for example, in the usual spherically
symmetric collapsing star the intersection of the event horizon
with  level sets of a time function will be a point at the time of
appearance of the event horizon.  We refer the reader
to~\cite[Section~3]{ChDGH} for other such examples, including one
in which the topology of sections of horizon   changes type from
toroidal to spherical as time evolves. This behaviour can be
traced back to the existence of past end points of the generators
of the horizon. Nevertheless, some sections of the horizon have
controlled topology -- for instance, we have the following:

\begin{ptctheorem}\label{Tqlbhtopo} Under the hypotheses of Theorem~\ref{Tqlbh1}, consider
  a connected component $\mcT_\alpha$ of $\mcT$ such that
  $\mcE_{\mcT_\alpha}\ne\emptyset$. Let
  $$\mcC_\alpha(\tau):= \partial J^+(\mcT _\alpha(\tau))\;.$$
  If
  $\mcC_\alpha(\tau)\cap\mcE_{\mcT _\alpha}$ is a topological
  manifold, then each connected component thereof has spherical
  topology.\chindex{horizons!topology of}
\end{ptctheorem}
\begin{proof}
  Consider the open subset $\mcM_\tau$ of $\bmcM$ defined as
  $$\mcM_\tau:=I^+(\mcC_\alpha(\tau);\bcM)\cap
  I^-(\mcT_\alpha;\bcM)\subset
  \langle\langle\mcT_\alpha\rangle\rangle\;.$$
  We claim that $\left(\mcM_\tau, g|_{\mcM_\tau}\right)$ is globally
  hyperbolic: indeed, let $p,q \in \mcM_\tau$; global hyperbolicity of
  $\bmcM$ shows that $J^-(p;\bmcM)\cap J^+(q;\bmcM)$ is a compact
  subset of $\bmcM$, which is easily seen to be included in
  $\mcM_\tau$. It follows that $J^-(p;\mcM_\tau)\cap J^+(q;\mcM_\tau)$
  is compact, as desired. By the usual decomposition we thus have
  $$\mcM_\tau\approx \R\times\cS\;,$$
  where $\cS$ is a Cauchy
  hypersurface for $\mcM_\tau$. Applying Theorem~\ref{Tqlbh2} to the
  globally hyperbolic space-time $\mcM_\tau$ (which has a \wnc\
  boundary $\mcT_\alpha\cap\{t>\tau\}$) one finds that $\mcM_\tau$ is
  simply connected, and thus so is $\cS$. Since $\mcC_\alpha(\tau)$
  and $\mcE_\alpha$ are null hypersurfaces in $\mcM$, it is easily
  seen that the closure in $\bcM$ of the Cauchy surface
  $\{0\}\times\cS$ intersects $\mcE_\alpha$ precisely at
  $\mcC_\alpha(\tau)\cap\mcE_{\mcT _\alpha}$. It follows that $\cS$ is
  a compact, simply connected, three dimensional topological manifold
  with boundary, and a classical result~\cite[Lemma~4.9]{Hempel} shows
  that each connected component of $\partial\cS$ is a sphere. The
  result follows now from
  $\partial\cS\approx\mcC_\alpha(\tau)\cap\mcE_{\mcT _\alpha}$.\qed
\end{proof}
Yet another class of ``good sections" of $\mcE_\mcT$ can be
characterized\footnote{I am grateful to G.~Galloway for useful
discussions concerning this question, as well as many other points
presented in this section.} as follows: suppose that
$\langle\langle\mcT_\alpha\rangle\rangle\cap \hyp_\tau$ is a
submanifold with boundary of $\mcM$ which is, moreover, a retract
of $\langle\langle\mcT_\alpha\rangle\rangle$. Then
$\langle\langle\mcT_\alpha\rangle\rangle\cap \hyp_\tau$ is simply
connected by Theorem~\ref{Tqlbh2}, and spherical topology of all
boundary components of
$\langle\langle\mcT_\alpha\rangle\rangle\cap \hyp_\tau$ follows
again from~\cite[Lemma~4.9]{Hempel}. It is not clear whether there
always exist time functions $t$ such that the retract condition is
satisfied; similarly it is not clear that there always exist
$\tau$'s for which the conditions of Theorem~\ref{Tqlbhtopo} are
met for metrics which are not stationary (one would actually want
``a lot of $\tau$'s"). It would be of interest to understand this
better.

 We have
an area theorem for $\mcE_\mcT$: \chindex{area theorem}
\begin{ptctheorem}\label{Tqlbharea} Under the hypotheses of Theorem~\ref{Tqlbh1}, suppose
  further that $\mcE_\mcT\ne\emptyset$.  Let $\hyp_a$, $a=1,2$ be two
  achronal spacelike embedded hypersurfaces of $C^2$ differentiability
  class, set $S_a= \hyp_a\cap\mcE_\mcT$. Then:
     \begin{enumerate}
     \item The area of $S_a$ is well defined.
     \item If $$S_1\subset J^-(S_2)\ ,
       $$
       then the area of $S_2$ is larger than or equal to that of
       $S_1$.  (Moreover, this is true even if the area of $S_1$ is
       counted with multiplicity\footnote{See~\cite{ChDGH} for
         details.} of generators provided that $S_1\cap S_2 =
       \emptyset$.)
\end{enumerate}
\end{ptctheorem}

We note that point 1 is less trivial as it appears, because
horizons can be rather rough sets, and it requires a certain
amount of work to establish that claim.
\begin{proof}
  The result is obtained by a mixture of methods of~\cite{ChDGH} and
  of~\cite{Galloway:fitopology}, and proceeds by contradiction: assume
  that the Alexandrov divergence $\theta_\Al$ of $\mcE_\mcT$ is
  negative, and consider the $S_{\epsilon,\eta,\delta}$ deformation of
  the horizon as constructed in Proposition~4.1 of~\cite{ChDGH}, with
  parameters chosen so that $\theta_{\epsilon,\eta,\delta}<0$. Global
  hyperbolicity implies the existence of an achronal null geodesic
  from $S_{\epsilon,\eta,\delta}$ to some cut $\mcT(\tau)$ of $\mcT$.
  The geodesic can further be chosen to be ``extremal", in the sense
  that it meets $\mcT(t)$ for the smallest possible value of $t$ among
  all generators of the boundary of $J^+(S_{\epsilon,\eta,\delta})$
  meeting $\mcT$. The argument of the proof of Theorem~1
  of~\cite{Galloway:fitopology} shows that this is incompatible with
  the null energy condition and with weak null convexity of
  $\mcT(\tau)$. It follows that $\theta_\Al\ge 0$, and the result
  follows from \cite[Proposition~3.3 and Theorem~6.1]{ChDGH}. \qed
\end{proof}
It immediately follows from the proof above that, under the
hypotheses of Theorem~\ref{Tqlbh1}, the occurrence of twice
differentiable  future trapped (compact) surfaces implies the
presence of a black hole region. The same result holds for
semi-convex compact surfaces which are trapped in an Alexandrov
sense, that is, \eq{trsur} holds with $\lambda$ there defined in a
way which should be clear from the discussion following
Theorem~\ref{Tdiff}. It is, however, not known if the existence of
marginally trapped surfaces -- whether defined in a classical, or
Alexandrov, or a viscosity sense -- does signal the occurrence of
black hole; it would be of interest to settle that.

In summary, we have shown that the quasi-local black holes,
defined using \wnc\ timelike hypersurfaces, or boundaries, possess
several properties usually associated with the Scri-based black
holes, without the associated problems. We believe they provide a
reasonable alternative, well suited for numerical calculations. In
the next section we  address the question of numerically finding
the resulting horizons.

\subsection{Finding horizons}
\chindex{horizons!numerical finders} Consider the large coordinate
spheres $\cS_{\tau,R_1}$ of Section~\ref{Snbh}; under reasonable
restrictions on $\mcM$ one would expect that $\partial
J^-({\cS_{\tau,R_1}})$ converges as $\tau\to\infty$
--- {\em e.g.\/} in Hausdorff topology --- to the event horizon
$\event$ of \eq{nbh4}. Similarly, the achronal boundaries
$\partial J^-(\mcT(\tau))$ should sometimes converge to the event
horizon $\mcE_\mcT$. Whenever that is the case, the $\partial
J^-({\cS_{\tau,R_1}})$'s or the $\partial J^-(\mcT(\tau))$'s can
be taken as good approximations of the event horizon when $\tau$
is chosen to be large enough. This is of practical significance,
as it provides a way of approximately locating the horizon in
numerical simulations. Let us establish one such approximation
result, for the $\partial J^-(\mcT(\tau))$'s; clearly the
corresponding result for the $J^-({\cS_{\tau,R_1}})$'s follows by
specialization. We assume that $\bmcM$ is spatially compact and
globally hyperbolic, which seem rather natural assumptions in the
context of numerical simulations of space-times:

\begin{ptctheorem}\label{Tfh1}\chindex{horizons!convergence of}
  Consider a globally hyperbolic space-time $(\bmcM,g)$ with compact
  Cauchy hypersurfaces $\hyp_\tau$ and timelike boundary $\mcT$, and
  suppose that \bel{Tfh1.1} \mcB_\mcT\ne\emptyset\;.\ee Then for any
  $\tau$ we have
  $$\partial J^-(\mcT(\sigma))\cap \hyp_\tau\to_{\sigma\to\infty}
  \mcE_\mcT\cap \hyp_\tau\;,$$
  where the convergence is in Hausdorff
  distance.
\end{ptctheorem}

\begin{proof}
  Under the usual identification of $\mcM$ with
  $\R\times\hyp_0$, let $f$ be the graphing function of $\partial
  J^-(\mcT(\sigma))$ over the projection $\proj_2\mcE$ of $\mcE$ on
  $\hyp_0$; $\proj_2\mcE$ is an open subset of $\hyp_0$ by the
  invariance-of-the-domain theorem. If $p=(t,\vec x)$ is such that
  $t<f(\vec x)$, then there exists a causal curve from $p$ to $\mcT$,
  hence $p\in J^-(\mcT(\sigma))$ for some $\sigma$. It follows that
  $$\proj_2\mcE = \cup_\sigma \proj_2J^-(\mcT(\sigma))\;.$$
  Let
  $f_\sigma$ be the graphing function of $\partial J^-(\mcT(\sigma))$
  over $\proj_2J^-(\mcT(\sigma)) \subset\hyp_0$; clearly $$
  \proj_2J^-(\mcT(\sigma))\subset \proj_2J^-(\mcT(\sigma'))\qquad
  \sigma<\sigma'\;,$$
  which shows that $f$ and all the $f_{\sigma'}$'s
  are defined on $\proj_2J^-(\mcT(\sigma))$ for $\sigma'>\sigma $. The
  $f_\tau$'s are uniformly Lipschitz and bounded over any compact
  subset $K\subset\proj_2\mcE $ of $\hyp_0$, and hence for any
  sequence $\tau_i$ there exists a subsequence such that the
  $f_{\tau_{i_j}}$'s converge to a Lipschitz function $h$. Since
  $f_\tau\le f$ for all $\tau$ we have $h\le f|_K$. Suppose that there
  exists $\vec x$ such that $h(\vx)< f(\vx)$, then
  $p=\left(\left(h(\vx)+f(\vx)\right)/2,\vx\right)$ is in $J^-(\mcT)$,
  hence in $J^-(\mcT(\sigma))$ for some $\sigma$, so that
  $\left(h(\vx)+f(\vx)\right)/2 \le f_\sigma(\vec x)$ which is not
  possible; we thus have $h=f|_K$. It follows that the $f_\tau$'s
  converge pointwise to $f$. The $f_\tau$'s are monotonously
  increasing as $\tau $ increases, and Dini's theorem implies that the
  $f_\tau$'s converge uniformly to $f$ over any compact set. The
  result follows now by elementary considerations using the fact that
  $\mcE_\mcT\cap \hyp_\tau$ is compact, being the intersection of a
  closed set with a compact one. \qed
\end{proof}
Theorem~\ref{Tfh1} gives partial justification for the numerical
analysis of Aninos~\cite{Anninos:1995ay}, Libson {et
al.}~\cite{Libson:1996dk}, or Mass\'o {et
al.}~\cite{Masso:1998fi},\footnote{We note that the claim
in~\cite{Anninos:1995ay,Libson:1996dk}, that the horizon acts as
an attractor for backwards-directed null geodesics, does not seem
to be justified. In fact, such a statement is
coordinate-dependent: it is easy to introduce coordinate systems
in the Schwarzschild space-time where backwards-directed null
geodesics will be actually repelled by the horizon.} where the
event horizon $\event$ is taken to be
$$\event \approx \partial J^-({\cS_{\tau,R_1}})\;,\quad \mbox{$\tau$
as big as the computer simulation allows}\;.$$ 
In those works $\partial J^-({\cS_{\tau,R_1}})$ is further
numerically approximated as the solution of an eikonal equation,
which leads to difficulties at
past end points of the generators of $\partial
J^-({\cS_{\tau,R_1}})$. We conjecture that a more stable method of
locating objects such as $\partial J^-({\cS_{\tau,R_1}})$ is
provided by the following straightforward modification of the
Fermat-type variational principle presented in the Appendix below,
see \eq{tau}: let $\cP(x)$ be defined as in the paragraph
preceding \eq{tau}, with $\mcH_\sigma$ there replaced by
${\cS_{\tau,R_1}}$, and let $\hat\gamma$ be the null lifts of
paths $\gamma$ as defined in the paragraph preceding \eq{nulll}.
An argument similar to that of the proof of Corollary~\ref{C1}
shows that in globally hyperbolic space-times the set $\partial
J^-({\cS_{\tau,R_1}})$ is a graph of a function $f$ such that \be
\label{tau2} f(\vx) = \sup_{\hat\gamma \in \cP(x)}
t(\hat\gamma(a))\ . \ee Such a maximization over null lifts $\hat
\gamma$ of paths $\gamma$ which have image in a fixed spacelike
hypersurface should be easy to perform numerically by e.g.
Newton's method. The variational principle above automatically
takes care of the past end-points of the horizon.

\bigskip

\noindent {\bf Acknowledgements} The author is grateful to
 Y.~Choquet-Bruhat,
G.~Horowitz, S.~Husa, R.~Narayan, R.~Parentani, D.~Polarski and
H.~Reeves for bibliographical advice. Thanks are due to
H.~Friedrich and G.~Galloway for suggesting many improvements to a
previous version of this manuscript, and to R.~Geroch and
J.~Jezierski  for useful comments and discussions.

\appendix
\section*{Appendix A: Future horizons are semi-convex}
\label{Aconv}  A hypersurface $\mcH\subset M$ will be said to be
{\em future null geodesically ruled} if every point $p\in \mcH$
belongs to a future inextensible null geodesic $\Gamma \subset
\mcH$; those geodesics will be called {\em the generators} of
$\mcH$. We emphasize that the generators are allowed to have past
endpoints on $\mcH$, but no future endpoints.  {\em Past null
geodesically ruled} hypersurfaces are defined by changing the time
orientation. Examples of future geodesically ruled hypersurfaces
include past Cauchy horizons ${\cal
  D}^-(\Sigma)$ of achronal sets $\Sigma$ \cite[Theorem~5.12]{PenroseDiffTopo} and black hole event horizons $\partial
J^-(\scri^+)$~\cite[p.~312]{HE}.

Let $\mathrm{dim} M=n+1$ and suppose that $\cal O$ is a domain in
$\R^n$. Recall that a continuous function $f: \cal O\to \R$ is
called semi--convex if there exists a $C^2$ function $\phi: \cal
O\to \R$ such that $f+\phi$ is convex. We shall say that the graph
of $f$ is a semi--convex hypersurface if $f$ is semi--convex. A
hypersurface $\mcH$ in a manifold $M$ will be said semi--convex if
$\mcH$ can be covered by coordinate patches ${\cal U}_\alpha$ such
that $\mcH\cap {\cal U}_\alpha$ is a semi--convex  graph for each
$\alpha$.

Consider an achronal hypersurface $\cH\ne \emptyset$ in a globally
hyperbolic space--time $(M,g)$. Let $t$ be a time function on $M$
which induces a diffeomorphism of $M$ with $\R \times \Sigma$ in
the standard way \cite{GerochDoD,Seifert}, with the level sets
$\Sigma_\tau\equiv \{p| t(p)=\tau\}$ of $t$ being Cauchy surfaces.
As usual we identify $\Sigma_0$ with $\Sigma$, and in the
identification above the curves $\R\times \{q\}$, $q\in\Sigma$,
are integral curves of $\nabla t$. Define
\begin{equation}
  \label{sh}
  \sh=\{q\in \Sigma| \R\times \{q\}\ \mbox{\rm intersects}\ \cH \}\ .
\end{equation}
For $q\in\sh$ the set $(I\times {q} )\cap H$ is a point by
achronality of $\cH$, we shall denote this point by $(f(q),q)$.
Thus an achronal hypersurface $\cH$ in a globally hyperbolic
space--time is a graph over $\sh$ of a function $f$. The
invariance-of-the-domain theorem shows that $\sh$ is an open
subset of $\Sigma$. We have the following:
\begin{aptctheorem}
\label{T1}\chindex{horizons!semi-convexity of} Let $\mcH\ne
\emptyset$ be an achronal future null geodesically ruled
hypersurface in a globally hyperbolic space--time
$(M=\R\times\Sigma,g)$. Then $\mcH$ is the graph of a semi--convex
function $f$ defined on an open subset $\sh$ of $\Sigma$, in
particular $\cH$ is semi--convex.
\end{aptctheorem}
\begin{proof}
  As discussed above, $\mcH$ is the graph of a function $f$.  The idea
  of the proof is to show that $f$ satisfies a variational principle,
  the semi--concavity of $f$ follows then by a standard argument. Let
  $p\in \mcH$ and let $\cal O$ be a coordinate patch in a neighborhood
  of $p$ such that $x^0 = t$, with $\cal O$ of the form $I\times
  B(3R)$, where $B(R)$ denotes a coordinate ball centered at $0$ of
  radius $R$ in $\R^3$, with $p=(t(p), 0)$. Here $I$ is the range of
  the coordinate $x^0$, we require it to be a bounded interval the
  size of which will be determined later on. We further assume that
  the curves $I\times \{\vx\}$, $\vx\in B(3R)$, are integral curves of
  $\nabla t$. Define
  $$
  {\cal U}_0=\{\vec x \in B(3R)| \mathrm{\ the\ causal\ path }\
  I\ni t \to (t,\vec x) \ \mathrm{intersects }\ \cH\}\ .
  $$
  We note that ${\cal U}_0$ is non--empty, since $0\in {\cal U}_0$.
  Set \be\label{Hsection} \cH_\sigma=\cH\cap \Sigma_\sigma \ , \ee and
  choose $\sigma$ large enough so that ${\cal O}\subset I^-(
  \Sigma_\sigma)$.  Now $p$ lies on a future inextensible generator
  $\Gamma$ of $\cH$, and global hyperbolicity of $(M,g)$ implies that
  $\Gamma\cap \Sigma_\sigma $ is nonempty, hence $\cH_\sigma$ is
  nonempty.

  For $\vx\in B(3R)$ let $\cP(x)$ denote the collection of piecewise
  differentiable future directed null curves $\Gamma: [a,b]\to M$ with
  $\Gamma(a)\in \R\times \{\vx\}$ and $\Gamma(b)\in \cH_\sigma$. We
  define \be \label{tau} \tau(\vx) = \sup_{\Gamma \in \cP(x)}
  t(\Gamma(a))\ . \ee We emphasize that we allow the domain of
  definition $[a,b]$ to depend upon $\Gamma$, and that the ``$a$"
  occurring in $ t(\Gamma(a))$ in \eq{tau} is the lower bound for the
  domain of definition of the curve $\Gamma$ under consideration. We
  have the following result (compare \cite{AP,GMP,Perlick}):

\begin{aptcproposition}[Fermat principle]\label{P1}\chindex{Fermat's
    principle} For $\vx\in\cU_0$ we have
  $$
  \tau(\vx)=f(\vx)\ .$$
\end{aptcproposition}
\begin{proof}
  Let $\Gamma$ be any generator of $\cH$ such that
  $\Gamma(0)=(f(\vx),\vx)$, clearly $\Gamma\in\cP(x)$ so that $
  \tau(\vx)\ge f(\vx)$. To show that this inequality has to be an
  equality, suppose for contradiction that $ \tau(\vx)>f(\vx)$, thus
  there exists a null future directed curve $\Gamma$ such that
  $t(\Gamma(0))> f(\vx)$ and $\Gamma(1)\in\cH_\sigma \subset \cH$.
  Then the curve $\tilde \Gamma$ obtained by following $\R\times
  \{\vx\}$ from $(f(\vx),\vx)$ to $(t(\Gamma(0)),\vx)$ and following
  $\Gamma$ from there on is a causal curve with endpoints on $\cH$
  which is not a null geodesic. By \cite[Proposition 4.5.10]{HE}
  $\tilde \Gamma$ can be deformed to a timelike curve with the same
  endpoints,
  which is impossible by achronality of $\cH$. \qed
\end{proof}
The Fermat principle, Proposition \ref{P1}, shows that $f$ is a
solution of the variational principle \eq{tau}. Now this
variational principle can be rewritten in a somewhat more
convenient form as follows: The identification of $M$ with
$\R\times\Sigma$ by flowing from $\Sigma_0\equiv \Sigma$ along the
gradient of $t$ leads to a global decomposition of the metric of
the form
$$
g = \alpha(-dt^2 + h_t)\ , $$ where $h_t$ denotes a $t$--dependent
family of Riemannian metrics on $\Sigma$. Any future directed
differentiable null curve $\Gamma(s)=(t(s),\vg(s))$ satisfies
$$
\frac{dt(
s)
}{ds} 
=\sqrt{h_{t(s)}(\dg,\dg)}\ , $$ where $\dg$ is a shorthand for
$d\vg(s)/ds$. It follows that for any $\Gamma\in\cP(x)$ it holds
that
\begin{eqnarray*}
  t(\Gamma(a)) & = & t(\Gamma(b)) - \int _a ^b \frac{dt}{ds} ds \\ & =
  & 
  \sigma- \int _a ^b \sqrt{h_\tgs(\dg,\dg)} ds\ .
\end{eqnarray*}
This allows us to rewrite \eq{tau} as \be \label{taun} \tau(\vx) =
\sigma - \mu(\vx)\ , \qquad \mu(\vx)\equiv\inf_{\Gamma \in
  \cP(x)} \int _a ^b \sqrt{h_{t(s)}(\dg,\dg)} ds\ .  \ee We note that
in static space--times $\mu(\vx)$ is the Riemannian distance from
$\vx$ to $\cH_\sigma$. In particular Equation \eq{taun} implies
the well known fact, that in globally hyperbolic static
space--times Cauchy horizons of open subsets of level sets of $t$
are graphs of the distance function from the boundary of those
sets.

Let $\vg:[a,b]\to \Sigma$ be a piecewise differentiable path, for
any $p\in \R\times \{\vg(b)\}$ we can find a null future directed
curve $\hat\gamma :[a,b] \to M$ of the form $\hat\gamma(s)
=(\phi(s),\vg(s))$ with future end point $p$ by solving the
problem
\be \cases{ \phi(b)=t(p)\ , & \cr
  \displaystyle{\frac{d\phi(s)}{ds}}=\sqrt{h_{\phi(s)}(\dg(s),\dg(s))}\
  . & } \label{nulll}\ee
  %
  %
  %
  %
The path $\hat\gamma$ will be called the {\em null lift of
$\gamma$ with endpoint $p$}.

As an example of application of Proposition \ref{P1} we recover
the following well known result \cite{PenroseDiffTopo}:

\begin{aptccorollary}\label{C1} $f$ is Lipschitz continuous
  on any compact subset of its domain of definition.
\end{aptccorollary}
\begin{proof}
  For $\vec y,\vec z\in B(2R)$ let $K\subset \R\times B(2R)$ be a
  compact set which contains all the null lifts $\Gamma_{\vy, \vz}$ of
  the coordinate segments $[\vec y,\vec z]:=\{\lambda \vec y
  +(1-\lambda)\vec z\ , \ \lambda \in [0,1]\}$ with endpoints
  $(\tau(\vec z),\vec z)$. Define \be \label{Cdef} \hat C = \sup
  \{\sqrt{h_p(n,n)}| p\in K, |n|_\delta =1\}\;, \ee where the supremum
  is taken over all points $p\in K$ and over all vectors $n\in T_pM$
  the coordinate components $n^i$ of which have Euclidean length
  $|n|_\delta$ equal to one.  Choose $I$ to be a bounded interval
  large enough so that $K\subset I\times B(2R)$ and, as before, choose
  $\sigma$ large enough so that $I\times B(2R)$ lies to the past of
  $\Sigma_\sigma$.  Let $\vec y,\vec z\in B(2R)$ and consider the
  causal curve $\Gamma=(t(s),\gamma(s))$ obtained by following the
  null lift $\Gamma_{\vy, \vz}$ in the parameter interval $s\in[0,1]$,
  and then a generator of $\cH$ from $(\tau(\vec z),\vec z)$ until
  $\cH_\sigma$ in the parameter interval $s\in[1,2]$. Then we have
  $$
  \mu(\vz)=\int_1^2 \sqrt{h_\tgs(\dg,\dG)} ds\;. $$
  Further
  $\Gamma\in \cP(x)$ so that
\begin{eqnarray}
  \mu(\vy) & \le & \int_0^2 \sqrt{h_\tgs(\dg,\dg)}
  ds\nonumber \\ & = & \int_0^1 \sqrt{h_\tgs(\dg,\dg)} ds +
  \int_1^2 \sqrt{h_\tgs(\dg,\dg)} ds \nonumber\\ & \le &
  \hat C |\vy-\vz|_\delta + \mu(\vz) \label{tin} \ , \end{eqnarray}
where $|\cdot|_\delta$ denotes the Euclidean norm of a vector, and
with $\hat C$ defined in \eq{Cdef}. Setting 1) first $\vy = \vx$,
$\vz = \vx + \vh$ in \eq{tin} and 2) then $\vz = \vx$, $\vy = \vx
+ \vh$, the Lipschitz continuity of $f$ on $B(2R)$ follows.  The
general result is obtained now by a standard covering argument.
\qed
\end{proof}
Returning to the proof of Theorem \ref{T1}, for $\vx\in B(R)$
  let $\Gamma_{\vx}$ 
  be a generator of $\cH$ such that $\Gamma_{\vx}(0)=(\tau(\vx),\vx)$,
  and, if we write $\Gamma_{\vx}(s)= (\phi_{\vx}(s),\gamma_{\vx}(s))$,
  then we require that $\gamma_{\vx}(s)\in B(2R) $ for $s\in[0,1]$.
  For $s\in[0,1]$ and $\vh\in B(R)$ let $\gamma_{\vx,\pm}(s)\in\Sigma$
  be defined by
  $$\gamma_{\vx,\pm}(s) = \gamma_{\vx}(s)\pm (1-s)\vh = s
  \gamma_{\vx}(s)+(1-s)(\gamma_{\vx}(s)\pm \vh)\in B(2R)\ . $$
  We note
  that
  $$\gamma_{\vx,\pm}(0)=\vx \pm \vh\ , \qquad \gamma_{\vx,\pm}(1)=
  \gamma_{\vx}(1)\ , \qquad \dot \gamma_{\vx,\pm} - \dot \gamma_{\vx}=
  \mp \vh\ . $$
  Let $\Gamma_{\vx,\pm}=(\pxpm,\gamma_{\vx,\pm})$ be the
  null lifts of the paths $\gamma_{\vx,\pm}$ with endpoints
  $\Gamma_{\vx}(1)$. Let $K$ be a compact set containing all the
  $\Gamma_{\vx,\pm}$'s, where $\vx$ and $\vh$ run through $ B(R)$. Let
  I be any bounded interval such that $I\times B(2R)$ contains $K$. As
  before, choose $\sigma$ so that $I\times B(2R)$ lies to the past of
  $\Sigma_\sigma$, and let $b$ be such that $\Gamma_{\vx}(b)\in
  \cH_\sigma$. (The value of the parameter $b$ will of course depend
  upon $\vx$). Let $\Gamma_\pm$ be the null curve obtained by
  following $\Gamma_{\vx,\pm}$ for parameter values $s\in [0,1]$, and
  then $ \Gamma_{\vx}$ for parameter values $s\in [1,b]$.  Then
  $\Gamma_\pm\in {\cP}(\vx\pm \vh)$ so that we have
  $$
  \mu(\vx\pm \vh)\le \int_0^1 \sqrt{h_\pxpm(\dot
    \gamma_{\vx,\pm},\dot \gamma_{\vx,\pm})} ds + \int_1^b
  \sqrt{h_\pxpm(\dot \gamma_{\vx},\dot \gamma_{\vx})} ds\ .
  $$
  Further
  $$\mu(\vx)= \int_0^b \sqrt{h_\pxhere(\dot \gamma_{\vx},\dot
    \gamma_{\vx})} ds\ , $$
  hence \begin{eqnarray} \lefteqn{
      \frac{\mu(\vx+ \vh)+ \mu(\vx- \vh)}{2} - \mu(\vx) \le }&&
    \nonumber \\ &&\int_0^1 \left(\frac{\sqrt{h_\pxpm(\dot
          \gamma_{\vx,+},\dot \gamma_{\vx,+})} + \sqrt{h_\pxpm(\dot
          \gamma_{\vx,-},\dot \gamma_{\vx,-})}}{2} -
      \sqrt{h_\pxhere(\dot \gamma_{\vx},\dot \gamma_{\vx})} \right)ds
    \; . \nonumber\\ && \label{concineq} \end{eqnarray} Since
  solutions of ODE's with parameters are differentiable functions of
  those, we can write
\begin{equation}
  \label{star}
  \phi _\pm(s)=\pxhere+\psi_i(s)h^i +r(s,h), \qquad |r(s,h)| \le C
|h|_\delta^2\ ,
\end{equation}
for some functions $\psi_i$, with a constant $C$ which is
independent of $\vx,\vh\in B(R)$ and $s\in [0,1]$. Inserting
\eq{star} in \eq{concineq}, second order Taylor expanding the
function $\sqrt{h_{\phi _\pm(s)}(\dot \gamma_{\vx,\pm},\dot
  \gamma_{\vx,\pm})}(s)$ in all its arguments around
$(\pxhere,\gamma_{\vx}(s),\dot \gamma_{\vx}(s))$ and using
compactness of $K$ one obtains \be \label{concineq2}
\frac{\mu(\vx+ \vh)+ \mu(\vx- \vh)}{2} - \mu(\vx) \le C
|\vh|_\delta^2 \ , \ee for some constant $C$. Set
$$
\psi(\vx)=\mu(\vx) - C |\vx|_\delta^2 \ . $$ Equation
\eq{concineq2} shows that
$$\forall \vx, \vh \in B(R) \qquad \psi(\vx)\ge \frac{\psi(\vx+ \vh)+
  \psi(\vx- \vh)}{2} \ . $$
A standard argument implies that $\psi$ is concave. It follows
that $$f(\vx)+ C |\vx|_\delta^2=\tau(\vx)+ C
|\vx|_\delta^2=\sigma-\mu(\vx)+ C
|\vx|_\delta^2=\sigma-\psi(\vx)$$ is convex, which is what had to
be established. \qed
\end{proof}

\def\cprime{$'$}
\providecommand{\bysame}{\leavevmode\hbox
to3em{\hrulefill}\thinspace}
\providecommand{\MR}{\relax\ifhmode\unskip\space\fi MR }
\providecommand{\MRhref}[2]{%
  \href{http://www.ams.org/mathscinet-getitem?mr=#1}{#2}
} \providecommand{\href}[2]{#2}

\index{black holes|)}


\printindex

\begin{thebibliography}{100}

\bibitem{Alcubierre:2001vm}
M.~Alcubierre, B.~Brugmann, D.~Pollney, E.~Seidel, and
T.~Takahashi,
  \emph{Black hole excision for dynamic black holes}, Phys.\ Rev. \textbf{D64}
  (2001), 061501 (5pp.), gr-qc/0104020.

\bibitem{Alcubierre:2000ke}
M.~Alcubierre et~al., \emph{The 3d grazing collision of two black
holes},
  (2000), gr-qc/0012079.

\bibitem{Alcubierre:1998rq}
\bysame, \emph{Test-beds and applications for apparent horizon
finders in
  numerical relativity}, Class. Quant. Grav. \textbf{17} (2000), 2159--2190,
  gr-qc/9809004.

\bibitem{manderson:static}
M.T. Anderson, \emph{On the structure of solutions to the static
vacuum
  {E}instein equations}, Annales H.~ Poincar\'e \textbf{1} (2000), 995--1042,
  gr-qc/0001018.

\bibitem{PRLetter}
L.~Andersson and P.T. Chru\'sciel, \emph{On ``hyperboloidal''
{C}auchy data for
  vacuum {E}instein equations and obstructions to smoothness of null infinity},
  Phys. Rev. Lett. \textbf{70} (1993), 2829--2832.

\bibitem{AndChDiss}
\bysame, \emph{On asymptotic behaviour of solutions of the
constraint equations
  in general relativity with ``hyperboloidal boundary conditions''}, Dissert.
  Math. \textbf{355} (1996), 1--100.

\bibitem{Anninos:1995ay}
P~Anninos, D.~Bernstein, S.~Brandt, J.~Libson, J.~Masso,
E.~Seidel, L.~Smarr,
  W.-M. Suen, and P.~Walker, \emph{Dynamics of apparent and event horizons},
  Phys. Rev. Lett. \textbf{74} (1995), 630--633, gr-qc/9403011.

\bibitem{AP}
F.~Antonacci and P.~Piccione, \emph{A {F}ermat principle on
{L}orentzian
  manifolds and applications}, Appl.\ Math.\ Lett. \textbf{9} (1996), 91--95.

\bibitem{Barcelo:2001ah}
C.~Barcel\'o, S.~Liberati, and M.~Visser, \emph{Analog gravity
from field
  theory normal modes?}, Class. Quant. Grav. \textbf{18} (2001), 3595--3610,
  gr-qc/0104001.

\bibitem{Barcelo:2001ca}
\bysame, \emph{Towards the observation of {Hawking radiation in
Bose-Einstein}
  condensates},  (2001), gr-qc/0110036.

\bibitem{BartnikChrusciel}
R.~Bartnik and P.T. Chru\'sciel, \emph{Spectral boundary
conditions for
  {D}irac--type equations}, in preparation.

\bibitem{Bartnik:NQSURL}
R.~Bartnik and A.~Norton, \emph{{NQS Einstein solver}},
  \url{http://beth.ise.canberra.edu.au/mathstat/StaffPages/Robert2.htm}.

\bibitem{Beem-Ehrlich:Lorentz}
J.~K. Beem and P.~E. Ehrlich, \emph{Global {Lorentzian} geometry},
Pure and
  Applied Mathematics, vol.~67, Marcel Dekker, New York, 1981.

\bibitem{BK2}
J.K. Beem and A.~Kr\'{o}lak, \emph{Cauchy horizon endpoints and
  differentiability}, Jour.\ Math.\ Phys. \textbf{39} (1998), 6001--6010,
  gr-qc/9709046.

\bibitem{BeigSimon3}
R.~Beig and W.~Simon, \emph{On the uniqueness of static
perfect--fluid
  solutions in general relativity}, Commun.\ Math.\ Phys. \textbf{144} (1992),
  373--390.

\bibitem{Bishop:1997ik}
N.T. Bishop, R.~Gomez, L.~Lehner, M.~Maharaj, and J.~Winicour,
  \emph{High-powered gravitational news}, Phys. Rev. \textbf{D56} (1997),
  6298--6309, gr-qc/9708065.

\bibitem{Boucher}
W.~Boucher, \emph{Cosmic no hair theorems}, Classical general
relativity (W.B.
  Bonnor and M.A.H. MacCallum, eds.), Cambridge University Press, Cambridge,
  1984, pp.~43--52.

\bibitem{BGH}
W.~Boucher, G.W. Gibbons, and G.T. Horowitz, \emph{Uniqueness
theorem for
  anti--de {S}itter spacetime}, Phys.\ Rev.\ D \textbf{30} (1984), 2447--2451.

\bibitem{Brandt:2000yp}
Steve Brandt et~al., \emph{Grazing collisions of black holes via
the excision
  of singularities}, Phys. Rev. Lett. \textbf{85} (2000), 5496--5499,
  gr-qc/0009047.

\bibitem{Bromley}
B.~Bromley, \emph{A lightweight review of middleweight black
holes}, Matters of
  gravity \textbf{14} (1999), 6--7, gr-qc/9909022.

\bibitem{Brout:1995rd}
R.~Brout, S.~Massar, R.~Parentani, and P.~Spindel, \emph{A primer
for black
  hole quantum physics}, Phys.\ Rept. \textbf{260} (1995), 329--454.

\bibitem{GallowayBrowdy}
S.F. Browdy and G.J. Galloway, \emph{{Topological censorship and
the topology
  of black holes.}}, Jour.\ Math.\ Phys. \textbf{36} (1995), 4952--4961.

\bibitem{BKK}
R.~Budzy\'nski, W.~Kondracki, and A.~Kr\'{o}lak, \emph{On the
differentiability
  of {C}auchy horizons}, Jour.\ Math.\ Phys. \textbf{40} (1999), 5138--5142.

\bibitem{bunting:masood}
G.~Bunting and A.K.M. Masood{--ul--A}lam, \emph{Nonexistence of
multiple black
  holes in asymptotically euclidean static vacuum space-time}, Gen.\ Rel.\
  Grav. \textbf{19} (1987), 147--154.

\bibitem{Carr}
B.J. Carr, \emph{Primordial black holes as a probe of the early
universe and a
  varying gravitational constant}, astro-ph/0102390.

\bibitem{Carter:1997im}
B.~Carter, \emph{Has the black hole equilibrium problem been
solved?}, in Proc.
  of the 8th Marcel Grossmann Meeting on Relativistic Astrophysics - MG 8,
  Jerusalem, Israel, 22 - 27 June 1997, T.~Piran, Ed., World Scientific, 1999,
  gr-qc/9712038.

\bibitem{CMSciama}
A.~Celotti, J.C. Miller, and D.W. Sciama, \emph{{Astrophysical
evidence for the
  existence of black holes}}, Class.\ Quantum Grav. \textbf{16} (1999),
  A3--A21, astro-ph/9912186.

\bibitem{CGME}
J.~Chaname, A.~Gould, and J.~Miralda-Escude, \emph{Microlensing by
the cluster
  of black holes around {Sgr A$^*$}},  (2001), astro-ph/0102481.

\bibitem{demetrios:scalar10}
D.~Christodoulou, \emph{Examples of naked singularity formation in
the
  gravitational collapse of a scalar field}, Ann. Math. \textbf{140} (1994),
  607--653.

\bibitem{Christodoulou:action}
\bysame, \emph{The action principle and partial differential
equations}, Annals
  of Mathematics Studies, vol. 146, Princeton Univ.\ Press, Princeton, 2000.

\bibitem{Ch-Kl}
D.~Christodoulou and S.~Klainermann, \emph{Nonlinear stability of
{M}inkowski
  space}, Princeton University Press, Princeton, 1993.

\bibitem{ChRT2}
P.T. Chru\'sciel, \emph{On the global structure of
{R}obinson--{T}rautman
  space--times}, Proc. Roy. Soc. London \textbf{A 436} (1992), 299--316.

\bibitem{Chnohair}
\bysame, \emph{``{N}o {H}air'' {T}heorems -- folklore,
conjectures, results},
  Differential Geometry and Mathematical Physics (J.~Beem and K.L. Duggal,
  eds.), Cont.\ Math., vol. 170, AMS, Providence, 1994, pp.~23--49,
  gr--qc/9402032.

\bibitem{ChAscona}
\bysame, \emph{Uniqueness of black holes revisited}, vol.~69,
Helv. Phys. Acta,
  May 1996, Proceedings of Journ\'es Relativistes 1996, gr-qc/9610010,
  pp.~529--552.

\bibitem{Chendpoints}
\bysame, \emph{A remark on differentiability of {C}auchy
horizons}, Class.\
  Quantum Grav. (1998), 3845--3848, gr-qc/9807059.

\bibitem{Chstatic}
\bysame, \emph{The classification of static vacuum space--times
containing an
  asymptotically flat spacelike hypersurface with compact interior}, Class.
  Quantum Grav. \textbf{16} (1999), 661--687, gr-qc/9809088.

\bibitem{Chstaticelvac}
\bysame, \emph{Towards the classification of static
electro--vacuum
  space--times containing an asymptotically flat spacelike hypersurface with
  compact interior}, Class. Quantum Grav. \textbf{16} (1999), 689--704,
  gr-qc/9810022.

\bibitem{ChDGH}
P.T. Chru\'sciel, E.~Delay, G.~Galloway, and R.~Howard,
\emph{Regularity of
  horizons and the area theorem}, Annales Henri Poincar\'e \textbf{2} (2001),
  109--178, gr-qc/0001003.

\bibitem{ChFGH}
P.T. Chru\'sciel, J.~Fu, G.~Galloway, and R.~Howard, \emph{On fine
  differentiability properties of horizons and applications to {R}iemannian
  geometry}, Jour.\ Geom.\ Phys. (2001), in press, gr-qc/0001003.

\bibitem{ChGalloway}
P.T. Chru\'sciel and G.J. Galloway, \emph{Horizons
non--differentiable on dense
  sets}, Commun.\ Math.\ Phys. \textbf{193} (1998), 449--470, gr-qc/9611032.

\bibitem{ChHerzlich}
P.T. Chru\'sciel and M.~Herzlich, \emph{The mass of asymptotically
hyperbolic
  {R}iemannian manifolds},  (2001), dg-ga/0110035.

\bibitem{ChImaxTaubNUT}
P.T. Chru\'sciel and J.~Isenberg, \emph{Non--isometric vacuum
extensions of
  vacuum maximal globally hyperbolic space--times}, Phys. Rev. \textbf{D48}
  (1993), 1616--1628.

\bibitem{CJK}
P.T. Chru\'sciel, J.~Jezierski, and J.~Kijowski,
\emph{{H}amiltonian field
  theory in the radiating regime}, Lect. Notes in Physics, vol. m70, Springer,
  Berlin, Heidelberg, New York, 2001, URL
  \url{http://www.phys.univ-tours.fr/~piotr/papers/hamiltonian_structure}.

\bibitem{ChMS}
P.T. Chru{\'s}ciel, M.A.H. MacCallum, and D.~Singleton,
\emph{Gravitational
  waves in general relativity. {XIV}: {B}ondi expansions and the
  ``polyhomogeneity'' of {S}cri}, Phil. Trans. Roy. Soc. London A \textbf{350}
  (1995), 113--141.

\bibitem{ChruscielSimon}
P.T. Chru\'sciel and W.~Simon, \emph{Towards the classification of
static
  vacuum spacetimes with negative cosmological constant}, Jour.\ Math.\ Phys.
  \textbf{42} (2001), 1779--1817, gr-qc/0004032.

\bibitem{ChWald}
P.T. Chru\'sciel and R.M. Wald, \emph{On the topology of
stationary black
  holes}, Class. Quantum Grav. \textbf{11} (1994), L147--152.

\bibitem{ClarkedeFelice2}
C.J.S. Clarke and F.~de~Felice, \emph{Globally noncausal
space-times}, Jour.\
  Phys.\ A \textbf{15} (1982), 2415--2417.

\bibitem{ColbertMushotzky}
E.J.M. Colbert and R.F. Mushotzky, \emph{The nature of accreting
black holes in
  nearby galaxy nuclei}, Astroph.\ Jour. \textbf{519} (1999), 89--107,
  astro-ph/9901023.

\bibitem{Dain:2001kn}
S.~Dain, \emph{Initial data for stationary space-times near
space-like
  infinity}, Class.\ Quantum Grav. \textbf{18} (2001), 4329--4338,
  gr-qc/0107018.

\bibitem{Damour:schmidt}
T.~Damour and B.~Schmidt, \emph{Reliability of perturbation theory
in general
  relativity}, Jour.\ Math.\ Phys. \textbf{31} (1990), 2441--2453.

\bibitem{Das:2000su}
S.R. Das and S.D. Mathur, \emph{The quantum physics of black
holes: Results
  from string theory}, Ann. Rev. Nucl. Part. Sci. \textbf{50} (2000), 153--206.

\bibitem{FlemingSoner}
W.H. Fleming and H.~Mete Soner, \emph{Controlled {M}arkov
processes and
  viscosity solutions}, Applications of Mathematics, vol.~25, Springer--Verlag,
  New York, Heidelberg, 1993.

\bibitem{friedrich:cauchy}
H.~Friedrich, \emph{Cauchy problem for the conformal vacuum field
equations in
  general relativity}, Commun.\ Math.\ Phys. \textbf{91} (1983), 445--472.

\bibitem{FriedrichdS}
\bysame, \emph{Existence and structure of past asymptotically
simple solutions
  of {E}instein's field equations with positive cosmological constant}, Jour.\
  Geom.\ Phys. \textbf{3} (1986), 101--117.

\bibitem{Friedrich:hyperbolicreview}
\bysame, \emph{Hyperbolic reductions for {E}instein's equations},
Class. and
  Quantum Grav. \textbf{13} (1996), 1451--1469.

\bibitem{Friedrich:Pune}
\bysame, \emph{Einstein's equation and geometric asymptotics},
{Gravitation and
  Relativity: At the turn of the Millennium, N.~Dahdich and J.~Narlikar
  (eds.)}, IUCAA, Pune, 1998, Proceedings of GR15, pp.~153--176.

\bibitem{Friedrich:2000qv}
H.~Friedrich and A.D. Rendall, \emph{The {Cauchy} problem for the
{E}instein
  equations}, Lect. Notes Phys. \textbf{540} (2000), 127--224, gr-qc/0002074.

\bibitem{FriedrichSchmidt}
H.~Friedrich and B.G. Schmidt, \emph{Conformal geodesics in
general
  relativity}, Proc.\ Roy.\ Soc.\ London Ser.\ A \textbf{414} (1987), 171--195.

\bibitem{galloway-topology}
G.J. Galloway, \emph{On the topology of the domain of outer
communication},
  Class. Quantum Grav. \textbf{12} (1995), L99--L101.

\bibitem{Galloway:fitopology}
\bysame, \emph{A ``finite infinity'' version of the {FSW}
topological
  censorship}, Class.\ Quantum Grav. \textbf{13} (1996), 1471--1478.

\bibitem{Galloway:1999br}
G.J. Galloway, K.~Schleich, D.~Witt, and E.~Woolgar, \emph{The
{AdS/CFT}
  correspondence conjecture and topological censorship}, Phys.\ Lett.
  \textbf{B505} (2001), 255--262, hep-th/9912119.

\bibitem{Galloway:2001uv}
G.J. Galloway, S.~Surya, and E.~Woolgar, \emph{A uniqueness
theorem for the
  {adS} soliton},  (2001), hep-th/0108170.

\bibitem{galloway:woolgar}
G.J. Galloway and E.~Woolgar, \emph{The cosmic censor forbids
naked topology},
  Class. Quantum Grav. \textbf{14} (1996), L1--L7, gr-qc/9609007.

\bibitem{GerochDoD}
R.~Geroch, \emph{Domain of dependence}, Jour. Math. Phys.
\textbf{11} (1970),
  437--449.

\bibitem{GerochEspositoWitten}
\bysame, \emph{Asymptotic structure of space-time}, Asymptotic
Structure of
  Space-Time (F.P. Esposito and L.~Witten, eds.), Plenum Press, New York, 1977,
  pp.~1--106.

\bibitem{GerochHorowitz}
R.~Geroch and G.~Horowitz, \emph{Asymptotically simple does not
imply
  asymptotically {M}inkowskian}, Phys.\ Rev.\ Lett. \textbf{40} (1978),
  203--206.

\bibitem{GKP}
R.~Geroch, E.H. Kronheimer, and R.~Penrose, \emph{{Ideal points in
  space-time.}}, Proc.\ Roy.\ Soc.\ London \textbf{A 327}, 545--567.

\bibitem{GMP}
F.~Giannoni, A.~Masiello, and P.~Piccione, \emph{A variational
theory for light
  rays in stably causal {L}orentzian manifolds: regularity and multiplicity
  results}, Commun.\ Math.\ Phys. \textbf{187} (1997), 375--415.

\bibitem{GibbonsGPI}
G.W. Gibbons, \emph{Gravitational entropy and the inverse mean
curvature flow},
  Class.\ Quantum Grav. \textbf{16} (1999), 1677--1687.

\bibitem{Gomez:1998uj}
R.~Gomez et~al., \emph{Stable characteristic evolution of generic
3-dimensional
  single-black-hole spacetimes}, Phys. Rev. Lett. \textbf{80} (1998),
  3915--3918, gr-qc/9801069.

\bibitem{Grandclement:2001ed}
P.~Grandcl\'ement, E.~Gourgoulhon, and S.~Bonazzola, \emph{Binary
black holes
  in circular orbits. {II. N}umerical methods and first results},  (2001),
  gr-qc/0106016.

\bibitem{Gundlach:1998us}
C.~Gundlach, \emph{Pseudo-spectral apparent horizon finders: An
efficient new
  algorithm}, Phys. Rev. \textbf{D57} (1998), 863--875, gr-qc/9707050.

\bibitem{HE}
S.W. Hawking and G.F.R. Ellis, \emph{The large scale structure of
space-time},
  Cambridge University Press, Cambridge, 1973.

\bibitem{Hempel}
J.~Hempel, \emph{3--manifolds}, Princeton University Press,
Princeton, 1976,
  Annals of Mathematics Studies No 86.

\bibitem{Heusler:book}
M.~Heusler, \emph{Black hole uniqueness theorems}, Cambridge
University Press,
  Cambridge, 1996.

\bibitem{Heusler:living}
\bysame, \emph{Stationary black holes: uniqueness and beyond},
Living Reviews
  \textbf{1} (1998), \url{http://www.livingreviews.org}.

\bibitem{Horowitz:1996rn}
G.T. Horowitz, \emph{Quantum states of black holes}, {R.M.~Wald
(ed.), Black
  holes and relativistic stars. Symposium held in Chicago, IL, USA, December
  14-15, 1996. Univ. of Chicago Press, Chicago, Ill., pp.~241--266}.

\bibitem{HorowitzMyers}
G.T. Horowitz and R.C. Myers, \emph{The {AdS/CFT} correspondence
and a new
  positive energy conjecture for general relativity}, Phys. Rev. \textbf{D59}
  (1999), 026005 (12 pp.).

\bibitem{HowardFu}
R.~Howard and J.H.G. Fu, \emph{Hypersurfaces with prescribed mean
curvature},
  (2001), preprint.

\bibitem{HI2}
G.~Huisken and T.~Ilmanen, \emph{The inverse mean curvature flow
and the
  {Riemannian P}enrose inequality}, Jour.\ Diff.\ Geom. (2002), in press.

\bibitem{Israel:vacuum}
W.~Israel, \emph{Event horizons in static vacuum space-times},
Phys. Rev.
  \textbf{164} (1967), 1776--1779.

\bibitem{Israel:bhreview}
\bysame, \emph{{Dark stars: The evolution of an idea}},
{S.W.~Hawking and
  W.~Israel (eds.), Three hundred years of gravitation. Cambridge: Cambridge
  University Press, pp.~199-276}, 1987.

\bibitem{Jacobson:1999zk}
T.~Jacobson, \emph{Trans-planckian redshifts and the substance of
the space-
  time river}, Prog.\ Theor.\ Phys.\ Suppl. \textbf{136} (1999), 1--17,
  hep-th/0001085.

\bibitem{Jacobson:venkatarami}
T.\ Jacobson and S.\ Venkatarami, \emph{Topology of event horizons
and
  topological censorship}, Class. Quantum Grav. \textbf{12} (1995), 1055--1061.

\bibitem{Kenney}
J.D.P. Kenney and E.E. Yale, \emph{Hubble space telescope imaging
of bipolar
  nuclear shells in the disturbed {Virgo Cluster Galaxy NGC 4438}},  (2000),
  astro-ph/0005052.

\bibitem{Kidder:2001tz}
L.E. Kidder, M.A. Scheel, and S.A. Teukolsky, \emph{Extending the
lifetime of
  3d black hole computations with a new hyperbolic system of evolution
  equations}, Phys. Rev. \textbf{D64} (2001), 064017, gr-qc/0105031.

\bibitem{KorGeb}
J.~Kormendy and K.~Gebhardt, \emph{Supermassive black holes in
galactic
  nuclei},  (2001), astro-ph/0105230.

\bibitem{KrieleHayward}
M.~Kriele and S.A. Hayward, \emph{{Outer trapped surfaces and
their apparent
  horizon}}, Jour.\ Math.\ Phys. \textbf{38}, 1593--1604.

\bibitem{Krolak:erpfsps}
A.~Kr{\'o}lak, \emph{The existence of regular partially future
asymptotically
  predictable space-times}, Jour.\ Math.\ Phys. \textbf{29} (1988), 1786--1788.

\bibitem{Kroon:detect}
J.A.V. Kroon, \emph{Can one detect a non-smooth null infinity?},
Class.\
  Quantum Grav. \textbf{18} (2001), 4311--4316.

\bibitem{Lafontaine}
J.~Lafontaine, \emph{Sur la g\'eom\'etrie d'une g\'en\'eralisation
de
  l'\'equation diff\'erentielle d'{O}bata}, Jour.\ de Math.\ Pures et Appli.
  \textbf{62} (1983), 63--72.

\bibitem{LafontaineRozoy}
J.~Lafontaine and L.~Rozoy, \emph{Courbure scalaire et trous
noirs},
  S\'eminaire de th\'eorie spectrale et g\'eom\'etrie, Grenoble \textbf{18}
  (2000), 69--76.

\bibitem{Lehner:2001wq}
L.~Lehner, \emph{Numerical relativity: A review}, Class.\ Quantum
Grav.
  \textbf{18} (2001), R25--R86, gr-qc/0106072.

\bibitem{Lengard}
O.~Lengard, \emph{{S}olutions of the {E}instein's equation, waves
maps, and
  semilinear waves in the radiation regime}, Ph.D. thesis, Universit\'e de
  Tours, 2001, \url{http://www/phys.univ-tours.fr/~piotr/papers/batz}.

\bibitem{Leonhardt:Piwnicki}
U.~Leonhardt and P.~Piwnicki, \emph{Relativistic effects of light
in moving
  media with extremely low group velocity}, Phys.\ Rev.\ Lett. \textbf{84}
  (2000), 822--825.

\bibitem{Leonhardt:2000hf}
\bysame, \emph{reply to the ``{C}omment on `{R}elativistic effects
of light in
  moving media with extremely low group velocity'{}'' by {M. V}isser}, Phys.
  Rev. Lett. \textbf{85} (2000), 5253, gr-qc/0003016.

\bibitem{Leray}
J.~Leray, \emph{Hyperbolic differential equations}, mimeographed
notes, 1953,
  Princeton.

\bibitem{Libson:1996dk}
J.~Libson, J.~Masso, E.~Seidel, W.-M. Suen, and P.~Walker,
\emph{Event horizons
  in numerical relativity. {I:M}ethods and tests}, Phys.\ Rev. \textbf{D53}
  (1996), 4335--4350, gr-qc/9412068.

\bibitem{Luminet}
J.-P. Luminet, \emph{{Black holes: A general introduction.}},
{F.W.~Hehl,
  C.~Kiefer and R.J.K.~Metzler (eds.), Black holes: theory and observation.
  Proceedings of the 179th W.E.~Heraeus seminar, Bad Honnef, Germany, August
  18-22, 1997. Springer Lect. Notes Phys. 514, pp.~3-34 }, astro-ph/9804006.

\bibitem{Macchetto}
F.D. Macchetto, \emph{Supermassive black holes and galaxy
morphology},  (2001),
  astro-ph/9910089.

\bibitem{Madejski}
G.M. Madejski, \emph{Black holes in active galactic nuclei:
observations},
  (1999), astro-ph/9903141.

\bibitem{Masso:1998fi}
J.~Masso, E.~Seidel, W.-M. Suen, and P.~Walker, \emph{Event
horizons in
  numerical relativity {II: A}nalyzing the horizon}, Phys.\ Rev. \textbf{D59}
  (1999), 064015 (19 pp.), gr-qc/9804059.

\bibitem{Narayan}
K.~Menou, E.~Quataert, and R.~Narayan, \emph{Astrophysical
evidence for black
  hole event horizons}, {Gravitation and Relativity: At the turn of the
  Millennium, N.~Dahdich and J.~Narlikar (eds.)}, IUCAA, Pune, 1998,
  Proceedings of GR15, pp.~43--65.

\bibitem{MerFer}
D.~Merritt and L.~Ferrarese, \emph{Relationship of black holes to
bulges},
  (2001), astro-ph/0107134.

\bibitem{MG}
J.~Miralda-Escud\'e and A.~Gould, \emph{A cluster of black holes
at the
  galactic center}, Astroph.\ Jour. \textbf{545} (2000), 847--853,
  astro-ph/0003269.

\bibitem{Misner}
C.W. Misner, \emph{Taub--{N}{U}{T} space as a counterexample to
almost
  anything}, Relativity Theory and Astrophysics, AMS, Providence, Rhode Island,
  1967, Lectures in Appl. Math., vol. 8, pp.~160--169.

\bibitem{MTW}
C.W. Misner, K.~Thorne, and J.A. Wheeler, \emph{Gravitation},
Freeman, San
  Fransisco, 1973.

\bibitem{MGH}
J.M. Morgan, L.J Greenhill, and J.R. Hernstein,
\emph{Observational evidence
  for massive black holes in the centers of active galaxies}, astro-ph/0002085.

\bibitem{NGM}
R.~Narayan, M.R. Garcia, and J.E. McClintock, \emph{X-ray {Novae}
and the
  evidence for black hole event horizons},  (2001), astro-ph/0107387.

\bibitem{Newman:coscen}
R.P.A.C. Newman, \emph{Cosmic censorship and curvature growth},
Gen.\ Rel.\
  Grav. \textbf{15} (1983), 641--353.

\bibitem{Novello:2001fv}
M.~Novello, J.M. Salim, V.A.~De Lorenci, and E.~Elbaz,
\emph{Nonlinear
  electrodynamics can generate a closed spacelike path for photons}, Phys.\
  Rev. \textbf{D63} (2001), 103516 (5 pp.).

\bibitem{ONeill}
B.~O'Neill, \emph{Semi--{R}iemannian geometry}, Academic Press,
New York, 1983.

\bibitem{Peet:2000hn}
A.W. Peet, \emph{Tasi lectures on black holes in string theory},
(2000),
  hep-th/0008241.

\bibitem{penrose:scri}
R.~Penrose, \emph{Zero rest-mass fields including gravitation},
Proc.\ Roy.\
  Soc.\ London \textbf{A284} (1965), 159--203.

\bibitem{PenroseDiffTopo}
R.~Penrose, \emph{Techniques of differential topology in
relativity}, SIAM,
  Philadelphia, 1972, (Regional Conf. Series in Appl. Math., vol. 7).

\bibitem{Perlick}
V.~Perlick, \emph{On {F}ermat's principle in general relativity:
I. {T}he
  general case}, Class.\ Quantum Grav. \textbf{7} (1990), 1319--1331.

\bibitem{PtakGriffiths}
A.~Ptak and R.~Griffiths, \emph{Hard {X}-ray variability in
{M~82}: Evidence
  for a nascent {AGN?}}, Astroph.\ Jour. \textbf{517} (1999), L85--L90,
  astro-ph/9903372.

\bibitem{Racke}
R.~Racke, \emph{Lectures on nonlinear evolution equations},
Friedr. Vieweg \&
  Sohn, Braunschweig/Wiesbaden, 1992.

\bibitem{Rees:Wald}
M.J. Rees, \emph{{Astrophysical evidence for black holes}},
{R.M.~Wald (ed.),
  Black holes and relativistic stars. Symposium held in Chicago, IL, USA,
  December 14-15, 1996. Univ. of Chicago Press, Chicago, Ill., pp.~79-101}.

\bibitem{ReesSupermassive}
\bysame, \emph{Supermassive black holes: Their formation, and
their prospects
  as probes of relativistic gravity}, Proceedings of ESO conference in honour
  of R.~Giacconi on Black Holes in Binaries and Galactic Nuclei, 2000,
  astro-ph/9912346.

\bibitem{Schmidtcqg91}
B.~Schmidt, \emph{On the uniqueness of boundaries at infinity of
asymptotically
  flat spacetimes}, Class.\ Quantum Grav. \textbf{8} (1991), 1491--1504.

\bibitem{Seifert}
H.J. Seifert, \emph{Smoothing and extending cosmic time
functions}, Gen. Rel.
  Grav. \textbf{8} (77), 815--831.

\bibitem{STScI}
The~Hubble site of~the {Space Telescope Science Institute},
  \url{http://hubble.stsci.edu}.

\bibitem{TamburinoWinicour}
L.A. Tamburino and J.H. Winicour, \emph{Gravitational fields in
finite and
  conformal {B}ondi frames}, Phys.\ Rev. \textbf{150} (1966), 1039--1053.

\bibitem{Taylor}
M.E. Taylor, \emph{Partial differential equations}, Springer, New
York, Berlin,
  Heidelberg, 1996.

\bibitem{harms}
Z.I. Tsvetanov, M.G. Allen, H.C. Ford, and R.J. Harms,
\emph{Morphology of the
  nuclear disk in {M87}}, 1998, Proceedings of the M87 Workshop, Ringberg
  castle, Germany, 15-19 Sep 1997, astro-ph/9803178.

\bibitem{Unruh}
W.G. Unruh, \emph{Dumb holes and the effects of high frequencies
on black hole
  evaporation}, Phys.\ Rev. \textbf{D51} (1995), 2827--2838, gr-qc/9409008.

\bibitem{Visser:mog}
M.~Visser, \emph{Optical black holes?}, Matters of gravity,
gr-qc/0002027.

\bibitem{Visser:2000pk}
\bysame, \emph{Comment on `{R}elativistic effects of light in
moving media with
  extremely low group velocity'}, Phys.\ Rev.\ Lett. \textbf{85} (2000), 5252,
  gr-qc/0002011.

\bibitem{Visser:1998qn}
Matt Visser, \emph{Acoustic black holes},  (1999), gr-qc/9901047.

\bibitem{Volovik:2000ua}
G.~E. Volovik, \emph{Superfluid analogies of cosmological
phenomena}, Phys.\
  Rept. \textbf{351} (2001), 195--348, gr-qc/0005091.

\bibitem{Wald:book}
R.M. Wald, \emph{General relativity}, University of Chicago Press,
Chicago,
  Ill., 1984.

\bibitem{Wald:LR}
\bysame, \emph{The thermodynamics of black holes}, Living Reviews
\textbf{4}
  (2001), gr-qc/9912119, \url{http://www.livingreviews.org/}.

\bibitem{Wang}
X.~Wang, \emph{Mass for asymptotically hyperbolic manifolds},
Jour.\ Diff.\
  Geom. \textbf{57} (2001), 273--299.

\bibitem{deZeeuw}
{T.~de} Zeeuw, \emph{Evidence for massive black holes in nearby
galactic
  nuclei},  (2000), astro-ph/0009249.

\bibitem{Zhang:hpet}
X.~Zhang, \emph{A definition of total energy-momenta and the
positive mass
  theorem on asymptotically hyperbolic 3 manifolds {I}},  (2001), preprint.

\end{thebibliography}
\end{document}